\documentclass[a4paper,11pt]{article}
\pdfoutput=1 

\usepackage{jcappub} 

\usepackage[T1]{fontenc} 

\usepackage{aasmacros}

\usepackage{amsmath}
\usepackage{bm}
\usepackage{tabularx}
\usepackage{mleftright}
\usepackage[normalem]{ulem}

\usepackage{rotating}

\usepackage{booktabs}

\usepackage{xcolor}
\definecolor{ao(english)}{rgb}{0.0, 0.5, 0.0}

\usepackage{soul} 

\newcommand{\mymap}[1]{\bm{#1}}
\newcommand{\myvector}[1]{\bar{#1}}
\newcommand{\mymatrix}[1]{\mathbf{#1}}

\newcommand{\ensemble}[1]{{\mleft<{#1}\mright>}}
\newcommand{\amplitudes}{$\mathcal{A}$}
\newcommand{\spectrals}{$\mathcal{B}$}

\usepackage{caption}
\usepackage{subcaption}

\makeatletter
\newcommand\notsotiny{\@setfontsize\notsotiny\@vipt\@viipt}
\makeatother

\title{\boldmath 
Determination of Polarization Angles in CMB Experiments and Application to CMB Component Separation Analyses}

\author[a,b]{E. de la Hoz\footnote{Corresponding author.},}
\author[a,b]{P. Diego-Palazuelos,}
\author[a]{E. Mart{\'\i}nez-González,}
\author[a]{P. Vielva,}
\author[a]{R. B. Barreiro}
\author[a,b]{J. D. Bilbao-Ahedo}

\affiliation[a]{Instituto de F\'isica de Cantabria (CSIC-Universidad de Cantabria),\\ 
Avda. de los Castros s/n, E-39005 Santander, Spain}
\affiliation[b]{Dpto. de F\'isica Moderna, Universidad de Cantabria, \\
Avda. los Castros s/n, E-39005 Santander, Spain}

\emailAdd{delahoz@ifca.unican.es}
\emailAdd{diegop@ifca.unican.es}
\emailAdd{martinez@ifca.unican.es}
\emailAdd{vielva@ifca.unican.es}
\emailAdd{barreiro@ifca.unican.es}
\emailAdd{bilbao@ifca.unican.es}

\abstract{The new generation of CMB polarization experiments will reach limits of sensitivity never achieved before in order to detect the elusive primordial $B$-mode signal. However, all these efforts will be futile if we lack a tight control of systematics. Here, we focus on the systematic that arises from the uncertainty on the calibration of polarization angles. Miscalibrated polarization angles induce a mixing of $E$- and $B$-modes that obscures the primordial $B$-mode signal. We introduce an iterative angular power spectra maximum likelihood-based method to calculate the polarization angles ($\hat{\alpha}$) from the multi-frequency signal. The basis behind this methodology grounds on nulling the $C_{\ell}^{EB}$ power spectra. In order  to simplify the likelihood, we assume that the rotation angles are small ($\lesssim 6^\circ$) and, the maximum likelihood solution for the rotation angles $\hat{\alpha}$, is obtained by applying an iterative process where the covariance matrix does not depend on $\hat{\alpha}$ per iteration, i.e., the rotation angles are fixed to the estimated $\hat{\alpha}$ in the previous iteration. With these assumptions, we obtain an analytical linear system which leads to a very fast computational implementation. We show that with this methodology we are able to determine the rotation angle for each frequency with sufficiently good accuracy. 
To prove the latter point we perform component separation analyses using the parametric component separation method B-SeCRET with two different approaches. 
In the first approach we apply the B-SeCRET pipeline to the signal de-rotated with the estimation of $\hat{\alpha}$, while in the second, the rotation angles are treated as model parameters using the estimation of $\hat{\alpha}$ as a prior information.
We obtain that the rotation angles estimations improve after applying the second approach, and show that the systematic residuals due to the non-null calibration polarization angles are mitigated to the order of a 1\% at the power spectrum level. }

\begin{document}
\maketitle
\flushbottom

\section{Introduction}
\label{sec:intro}

With temperature data successfully characterized thanks to the Planck satellite \cite{planck_2018}, the next generation of cosmic microwave background (CMB) experiments will be focused on the measurement of the CMB polarization. Amongst other scientific objectives \cite[see, e.g., ][]{simons_observatory, s4, pico}, the CMB polarization would allow us to further constrain the $\Lambda$-CDM cosmological model and the inflation paradigm, map the distribution of matter in the universe, and probe the density of light relics from (and beyond) the Standard Model of particle physics. To accomplish these science goals, future CMB experiments will need to meet very demanding sensitivity requirements, and, more importantly,  have a tight control on systematics.

One of the principal known sources of systematic uncertainty is the miscalibration of the polarization angle. A non-zero polarization angle of the polarimeter is one of the many systematics that lead to the mixing of the measured $E$- and $B$-modes, which, in turn, results in a leakage of $E$-modes into $B$-modes that can obscure the primordial $B$-mode signal from inflation \cite{hu2003,shimon2008,yadav2010,LiteBIRD_PTEP}. 
To reach their sensitivity requirements of $\sigma_r\sim10^{-3}$, future experiments can only afford uncertainties at the arcminute level on the determination of the polarization angle. Therefore, a very precise calibration is mandatory.

The common procedure for polarization angle calibration is to use an artificial calibrator (classically a thermal source behind a grid of linear polarizers), a natural astrophysical source of known polarization orientation (like, for example, the Crab Nebula \cite{calibration_with_crab}), or previous CMB measurements \cite{calibration_with_bicep}, to adjust the polarization angle of the detectors. Examples of experiments that rely on artificial calibration sources are SPTPol \cite{calibration_SPTPol} and BICEP \cite{calibration_BICEP}, while POLARBEAR \cite{calibration_POLARBEAR} and ACTPol \cite{calibration_ACTPol} use instead the Crab Nebula as their calibration reference. However, this type of calibration has its limitations \cite{keating2013}. On the one hand, artificial calibration sources have spectral properties that are very different from those of the CMB, and they cannot always be placed and/or maintained in the antenna's far-field during the whole observational campaign. However, new approaches has been propossed in the last years to place artificial calibration sources on stratospheric balloons~\cite{Nati2017}, CubeSat on a low-Earth-orbit~\cite{Johnson2015}, and on a satellite following a formation-flight with the CMB master satellite~\cite{casas2021}. On the other hand, the polarized emission of astrophysical sources needs to be very well constrained beforehand for it to serve as a calibration source, and even in that case, astrophysical sources suffer from time, frequency and spatial variability, and might not be visible from all CMB observatories. In addition, both astrophysical and artificial sources are much brighter than the CMB $B$-mode, which can make calibration more difficult by causing non-linearities in the response of the detectors. As a consequence, only a $\sim 0.5^\circ-1^\circ$ uncertainty can be reached with this kind of calibration methodologies \cite{calibration_ACTPol, calibration_BICEP, calibration_POLARBEAR, calibration_SPTPol, keating2013,calsat}. 

Another possibility is to use our current understanding of the universe and the measured CMB itself as a calibration reference. The existence of a primordial magnetic field at the last scattering surface, or a hypothetical parity-violating field or process, could produce the rotation of the CMB polarization plane \cite{pogosian2019,lue1999,primordial_magnetic_field}. This rotation, quantified by the cosmic birefringence angle, would also produce mixing between $E$- and $B$-modes. Nevertheless, no parity-violating processes are contemplated in either the $\Lambda$-CDM cosmological model or the Standard Model of particle physics, and no evidence of a significant primordial magnetic field has been detected (see \cite{ACTPol_birefringence, SPTPol_birefringence,gruppuso2020planck} for current constraints on cosmic birefringence), and, in the absence of such elements, the CMB is expected to have null $TB$ and $EB$ cross-correlations. In this way, if a zero birefringence angle is assumed, polarimeters can be calibrated by forcing the observed $TB$ and $EB$ angular power cross-spectra to be consistent with zero \cite{keating2013}. This methodology, known as self-calibration, is sometimes used in redundancy with other calibration methods to ensure consistency \cite{calibration_ACTPol, calibration_POLARBEAR}, and, depending on the noise level of the experiment at hand, it can lead to uncertainties around the arcminute scale \cite{keating2013}. 

However, the existence of parity-violating physics or a primordial magnetic field cannot be probed through CMB data when measurements are calibrated by imposing the cancellation of the $EB$ cross-correlation. To bypass this limitation, new algorithms have been proposed in recent years to simultaneously determine the uncertainty in the polarization angle and a possible non-zero $EB$ cross-correlation\footnote{In fact, an extension of the methodology proposed here which includes the determination of the birefringence angle will be presented in  \cite{birefringence_diego}.} \cite{minami2019simultaneous,yuto_partial-sky,yuto_cross-spectra}. A simultaneous determination of both the polarization and birefringence angles is possible as the Galactic foregrounds photons are unaffected by the cosmic birefringence due to their small propagation length \cite{minami2019simultaneous}. In this way, Galactic foreground emission can be used to break the degeneracy between cosmic birefringence and the polarization angle. 

Here we present a new implementation of the in-flight calibration methodology to determine the rotation angle from CMB polarization measurements, under the assumption of a null birefringence angle, presented in \cite{yuto_cross-spectra}. In our approach we introduce two approximations that lead to an analytical linear system of equations, which leads to an efficient and fast-converging algorithm to calculate rotation angles. Building on the results from this methodology, we apply two component separation approaches to remove this systematic from the cleaned CMB map. In the first approach we apply the B-SeCRET (Bayesian-Separation of Components and Residuals Estimate Tool) method \cite{de2020detection} to the signal after de-rotating with the estimation of the rotation angles. In the second approach, we incorporate the rotation angles into the component separation method itself as model parameters. Thus, the second approach constitutes an independent method to calculate the rotation angles. However, the information regarding the rotation angles from the previous method are used as prior information to help with convergence.

This work is structured as follows. A detailed explanation of our methodology  to estimate rotation angles is presented in section \ref{sec:methodology}. To test and validate our algorithm, we have produced sky simulations of a LiteBIRD-like experiment, which are described in section \ref{sec:simulations}. The results of applying our methodology to those simulations are then shown in section \ref{sec:performance}. The interplay between our methodology and component separation is presented in section \ref{sec:component_separation}. Final comments and conclusions are left for section \ref{sec:conclusions}.

\section{Methodology}
\label{sec:methodology}
Let $N$ be the number of frequency channels of the experiment, and $\alpha_{i}$ the $i$-th channel's rotation angle. Assuming a null birefringence angle, the spherical harmonics coefficients of the observed $E$ and $B$ polarization fields of the $i$-th channels are rotated as follows \cite{minami2019simultaneous}:
\begin{equation}
    \begin{pmatrix}
		E_{i,\ell,m} \\
	    B_{i,\ell,m}
	\end{pmatrix}
	=
	\begin{pmatrix}
		\mathrm{c}\mleft(2\alpha_i\mright) & -\mathrm{s}\mleft(2\alpha_i\mright)\\
	    \mathrm{s}\mleft(2\alpha_i\mright) & \mathrm{c}\mleft(2\alpha_i\mright)\\
	\end{pmatrix}
	\begin{pmatrix}
		\widetilde{E}_{i,\ell,m} \\
	    \widetilde{B}_{i,\ell,m}
	\end{pmatrix} \, ,
	\label{eq:EB_rotated}
\end{equation}
where $\widetilde{E}_{i,\ell,m}$ and $\widetilde{B}_{i,\ell,m}$ are the spherical harmonics coefficients of the non-rotated signal and, $\mathrm{c}()$ and $\mathrm{s}()$ stands for cosine and sine respectively. Thus, the observed angular power spectra are 
\begin{equation}
    \begin{pmatrix}
        C_{\ell}^{E_iE_j} \\
        C_{\ell}^{E_iB_j} \\
        C_{\ell}^{E_jB_i} \\
        C_{\ell}^{B_iB_j} 
    \end{pmatrix}
    =
    \begin{pmatrix}
        \mathrm{c}\mleft(2\alpha_i\mright)\mathrm{c}\mleft(2\alpha_j\mright)  & -\mathrm{c}\mleft(2\alpha_i\mright)\mathrm{s}\mleft(2\alpha_j\mright) &
        -\mathrm{s}\mleft(2\alpha_i\mright)\mathrm{c}\mleft(2\alpha_j\mright) & \mathrm{s}\mleft(2\alpha_i\mright)\mathrm{s}\mleft(2\alpha_j\mright) \\
        \mathrm{c}\mleft(2\alpha_i\mright)\mathrm{s}\mleft(2\alpha_j\mright)  & \mathrm{c}\mleft(2\alpha_i\mright)\mathrm{c}\mleft(2\alpha_j\mright) &
        -\mathrm{s}\mleft(2\alpha_i\mright)\mathrm{s}\mleft(2\alpha_j\mright) & -\mathrm{s}\mleft(2\alpha_i\mright)\mathrm{c}\mleft(2\alpha_j\mright) \\
        \mathrm{s}\mleft(2\alpha_i\mright)\mathrm{c}\mleft(2\alpha_j\mright)  & -\mathrm{s}\mleft(2\alpha_i\mright)\mathrm{s}\mleft(2\alpha_j\mright) &
        \mathrm{c}\mleft(2\alpha_i\mright)\mathrm{c}\mleft(2\alpha_j\mright) & -\mathrm{c}\mleft(2\alpha_i\mright)\mathrm{s}\mleft(2\alpha_j\mright) \\
        \mathrm{s}\mleft(2\alpha_i\mright)\mathrm{s}\mleft(2\alpha_j\mright)  & \mathrm{s}\mleft(2\alpha_i\mright)\mathrm{c}\mleft(2\alpha_j\mright) &
        \mathrm{c}\mleft(2\alpha_i\mright)\mathrm{s}\mleft(2\alpha_j\mright)  & \mathrm{c}\mleft(2\alpha_i\mright)\mathrm{c}\mleft(2\alpha_j\mright) 
    \end{pmatrix}
    \begin{pmatrix}
        C_{\ell}^{\widetilde{E}_i\widetilde{E}_j}\\
        C_{\ell}^{\widetilde{E}_i\widetilde{B}_j} \\
        C_{\ell}^{\widetilde{E}_j\widetilde{B}_i} \\
        C_{\ell}^{\widetilde{B}_i\widetilde{B}_j} 
    \end{pmatrix}
    \, .
    \nonumber
\end{equation}
The spectra can be combined to obtain the following relationships:
\begin{align}
    C_{\ell}^{E_iE_j} + C_{\ell}^{B_iB_j} & = \mleft(C_{\ell}^{\widetilde{E}_i\widetilde{E}_j} + C_{\ell}^{\widetilde{B}_i\widetilde{B}_j}\mright)\mathrm{c}\mleft(\phi\mright) + \mleft(C_{\ell}^{\widetilde{E}_i\widetilde{B}_j} - C_{\ell}^{\widetilde{E}_j\widetilde{B}_i}\mright)\mathrm{s}\mleft(\phi\mright) \, , \label{eq:sum_EE_BB}\\
    C_{\ell}^{E_iE_j} - C_{\ell}^{B_iB_j} & = \mleft(C_{\ell}^{\widetilde{E}_i\widetilde{E}_j} - C_{\ell}^{\widetilde{B}_i\widetilde{B}_j}\mright)\mathrm{c}\mleft(\psi\mright) - \mleft(C_{\ell}^{\widetilde{E}_i\widetilde{B}_j} + C_{\ell}^{\widetilde{E}_j\widetilde{B}_i}\mright)\mathrm{s}\mleft(\psi\mright) \, , \label{eq:minus_EE_BB}\\
    C_{\ell}^{E_iB_j} + C_{\ell}^{E_jB_i} & = \mleft(C_{\ell}^{\widetilde{E}_i\widetilde{E}_j} - C_{\ell}^{\widetilde{B}_i\widetilde{B}_j}\mright)\mathrm{s}\mleft(\psi\mright) + \mleft(C_{\ell}^{\widetilde{E}_i\widetilde{B}_j} + C_{\ell}^{\widetilde{E}_j\widetilde{B}_i}\mright)\mathrm{c}\mleft(\psi\mright) \, , \label{eq:sum_EB_BE}\\
    C_{\ell}^{E_iB_j} - C_{\ell}^{E_jB_i} & = -\mleft(C_{\ell}^{\widetilde{E}_i\widetilde{E}_j} + C_{\ell}^{\widetilde{B}_i\widetilde{B}_j}\mright)\mathrm{s}\mleft(\phi\mright) + \mleft(C_{\ell}^{\widetilde{E}_i\widetilde{B}_j} - C_{\ell}^{\widetilde{E}_j\widetilde{B}_i}\mright)\mathrm{c}\mleft(\phi\mright) \, , \label{eq:minus_EB_BE}
\end{align}
where $\psi = 2\mleft(\alpha_i+\alpha_j\mright)$ and $\phi = 2\mleft(\alpha_i-\alpha_j\mright)$. Substituting \eqref{eq:sum_EE_BB} and \eqref{eq:minus_EE_BB} in \eqref{eq:minus_EB_BE} and \eqref{eq:sum_EB_BE} respectively, and rearranging the terms we, get
\begin{align}
    \mleft(C_{\ell}^{E_iB_j} + C_{\ell}^{E_jB_i}\mright)\mathrm{c}\mleft(\psi\mright) & = \mleft(C_{\ell}^{E_iE_j} - C_{\ell}^{B_iB_j}\mright) \mathrm{s}\mleft(\psi\mright) + C_{\ell}^{\widetilde{E}_i\widetilde{B}_j} + C_{\ell}^{\widetilde{E}_j\widetilde{B}_i} \, , \label{eq:comb1}\\
    \mleft(C_{\ell}^{E_iB_j} - C_{\ell}^{E_jB_i}\mright)\mathrm{c}\mleft(\phi\mright) & = -\mleft(C_{\ell}^{E_iE_j} + C_{\ell}^{B_iB_j}\mright) \mathrm{s}\mleft(\phi\mright) + C_{\ell}^{\widetilde{E}_i\widetilde{B}_j} - C_{\ell}^{\widetilde{E}_j\widetilde{B}_i} \, . \label{eq:comb2}
\end{align}
Summing \eqref{eq:comb2} and \eqref{eq:comb1} and isolating the $C_{\ell}^{\widetilde{E}_i\widetilde{B}_{j}}$
\begin{align}
    C_{\ell}^{\widetilde{E}_i\widetilde{B}_{j}} =   \mleft(\mright. & \mleft. -\mathrm{c}\mleft(2\alpha_i\mright)\mathrm{s}\mleft(2\alpha_j\mright)C_{\ell}^{E_iE_j} + \mathrm{s}\mleft(2\alpha_i\mright)\mathrm{c}\mleft(2\alpha_j\mright)C_{\ell}^{B_iB_j} \mright. \nonumber \\
     & \mleft.  +\mathrm{c}\mleft(2\alpha_i\mright)\mathrm{c}\mleft(2\alpha_j\mright)C_{\ell}^{E_iB_j} - \mathrm{s}\mleft(2\alpha_i\mright)\mathrm{s}\mleft(2\alpha_j\mright)C_{\ell}^{E_jB_i} \mright) \, \label{eq:true_eb} .
\end{align}
Taking ensemble averages\footnote{Notice that $<C_{\ell}^{\widetilde{E}_i\widetilde{B}_j}>=<C_{\ell}^{\widetilde{E}_i\widetilde{B}_j,fg}> + <C_{\ell}^{\widetilde{E}_i\widetilde{B}_j,CMB}>$} we obtain 
\begin{align}
    \mleft<C_{\ell}^{E_iB_j}\mright>\mathrm{c}\mleft(2\alpha_i\mright)\mathrm{c}\mleft(2\alpha_j\mright) - &\mleft<C_{\ell}^{E_jB_i}\mright>\mathrm{s}\mleft(2\alpha_i\mright)\mathrm{s}\mleft(2\alpha_j\mright) = \nonumber\\ & \mleft<C_{\ell}^{E_iE_j}\mright>\mathrm{c}\mleft(2\alpha_i\mright)\mathrm{s}\mleft(2\alpha_j\mright)  - \mleft<C_{\ell}^{B_iB_j}\mright>\mathrm{s}\mleft(2\alpha_i\mright)\mathrm{c}\mleft(2\alpha_j\mright) \nonumber \\
    & + \mleft<C_{\ell}^{\widetilde{E}_i\widetilde{B}_j,fg}\mright> + \mleft<C_{\ell}^{\widetilde{E}_i\widetilde{B}_j,CMB}\mright>\, . \label{eq:ensemble_average}
\end{align}
We neglect the $ <C_{\ell}^{\widetilde{E}_i\widetilde{B}_j,CMB}>$ contribution as $\Lambda$-CDM predicts a null CMB $EB$ angular power spectrum. On the other hand, there is no theoretical model of the foregrounds' $EB$ spectrum. Current measurements are compatible with a statistically null foregrounds $EB$ angular power spectrum. Thus, in this work, we have assumed $<C_{\ell}^{\widetilde{E}_i\widetilde{B}_j,fg}> \approx 0$. Notice however, that this contribution could be taken into account with a proper model of the foregrounds' $EB$ power spectrum \cite{planck_EB_foregrounds1,planck_foregrounds_EB2}. Finally, adding from \eqref{eq:ensemble_average} the resulting equation from exchanging the $i$ and $j$ indices in \eqref{eq:ensemble_average} times $\mathrm{s}(2\alpha_i)\mathrm{s}(2\alpha_j)/(\mathrm{c}(2\alpha_i)\mathrm{c}(2\alpha_j))$ and isolating the observed cross EB spectrum we find
\begin{equation}
    C_{\ell}^{E_iB_j} = \dfrac{1}{\mathrm{c}\mleft(4\alpha_i\mright)+\mathrm{c}\mleft(4\alpha_j\mright)} \mleft(\mathrm{s}\mleft(4\alpha_j\mright)C_{\ell}^{E_iE_j} -\mathrm{s}\mleft(4\alpha_i\mright)C_{\ell}^{B_iB_j}\mright) \, .
    \label{eq:observed_EB}
\end{equation}
Thus, the log-likelihood\footnote{Notice that we are assuming no correlations among different multipoles since we saw that this assumption does not change the results.} (assumed to be Gaussian) of all the possible $EB$-spectra is given in \eqref{eq:full_likelihood}. The full log-likelihood contains an additional term proportional to the logarithm of the covariance determinant. In our methodology, this is assumed to be constant in each iteration, hence we can safely ignore it. However, this term is necessary in MCMC applications such as \cite{yuto_cross-spectra} as it is shown in \cite{birefringence_diego}.
\begin{equation}
    -2 \log \mathcal{L} \propto \sum\limits_{\ell=\ell_{m}}^{\ell_{M}} \mleft(\myvector{C}^{EB}_{\ell}-\dfrac{\mathrm{s}\mleft(4\myvector{\alpha}_1\mright)\myvector{C}_{\ell}^{EE}-\mathrm{s}\mleft(4\myvector{\alpha}_2\mright)\myvector{C}_{\ell}^{BB}}{\mathrm{c}\mleft(4\myvector{\alpha}_1\mright)+\mathrm{c}\mleft(4\myvector{\alpha}_2\mright)}\mright)^{T}\mymatrix{C}^{-1}_{\ell}\mleft(\myvector{C}^{EB}_{\ell}-\dfrac{\mathrm{s}\mleft(4\myvector{\alpha}_1\mright)\myvector{C}_{\ell}^{EE}-\mathrm{s}\mleft(4\myvector{\alpha}_2\mright)\myvector{C}_{\ell}^{BB}}{\mathrm{c}\mleft(4\myvector{\alpha}_1\mright)+\mathrm{c}\mleft(4\myvector{\alpha}_2\mright)}\mright) \, ,
    \label{eq:full_likelihood}
\end{equation}
where $\myvector{C}^{XY}_{\ell}$, $\myvector{\alpha}_1$ and $\myvector{\alpha}_2$ are $N^2$ vectors given by
\begin{align}
    \myvector{C}^{XY}_{\ell} & = \mleft(C^{X_1Y_1}_{\ell}\,\,  C^{X_1Y_2}_{\ell} \,\, ... \,\, C^{X_1Y_{N}}_{\ell} \,\, C^{X_2Y_1}_{\ell} \,\, ... \,\, C^{X_{2}Y_{N}}_{\ell} \,\,\, ... \,\,\, C^{X_{N}Y_{1}}_{\ell} \,\, ... \,\, C^{X_{N}Y_{N}}_{\ell} \mright)^{T} \, , \\
    \myvector{\alpha}_{1} & = \mleft(\alpha_1 \,\, \alpha_2 \,\, ... \,\, \alpha_N \,\, \alpha_1 ... \,\, \alpha_N \,\,\, ... \,\,\, \alpha_1 \,\, ... \,\, \alpha_N \mright)^{T} \, ,\\
    \myvector{\alpha}_{2} & = \mleft(\alpha_1 \,\, \alpha_1 \,\, ... \,\, \alpha_1 \,\, \alpha_2 \,\, ... \,\, \alpha_2 \,\,\, ... \,\,\, \alpha_N \,\, ... \,\, \alpha_N \mright)^{T} \, ,
\end{align}
$\ell_{m}$ and $\ell_{M}$ are the minimum and maximum multipoles involved in the likelihood, and $\mymatrix{C}_{\ell}$ is the $N^2\times N^2$ covariance matrix whose element $ijmn \equiv (i-1)N+j,(m-1)N+n$, i.e., the element at the $(i-1)N+j$ row and $(m-1)N+n$ column,  is given by
\begin{align}\label{eq:cov_matrix_cross}
     C_{\ell,ijmn}  = &\dfrac{1}{\mleft(2\ell+1\mright)}\mleft[\mleft<C_{\ell}^{E_iE_m}\mright>\mleft<C_{\ell}^{B_jB_n}\mright> + \mleft<C_{\ell}^{E_iB_n}\mright>\mleft<C_{\ell}^{E_mB_j}\mright>\mright. \\
     & + \dfrac{\mathrm{s}\mleft(4\alpha_m\mright)}{\mathrm{c}\mleft(4\alpha_m\mright)+\mathrm{c}\mleft(4\alpha_n\mright)} \mleft(\mleft<C_{\ell}^{E_iB_m}\mright>\mleft<C_{\ell}^{B_jB_n}\mright> + \mleft<C_{\ell}^{E_iB_n}\mright>\mleft<C_{\ell}^{B_mB_j}\mright>\mright) \nonumber\\
     & - \dfrac{\mathrm{s}\mleft(4\alpha_n\mright)}{\mathrm{c}\mleft(4\alpha_m\mright)+\mathrm{c}\mleft(4\alpha_n\mright)} \mleft(\mleft<C_{\ell}^{E_iE_m}\mright>\mleft<C_{\ell}^{B_jE_n}\mright> + \mleft<C_{\ell}^{E_iE_n}\mright>\mleft<C_{\ell}^{E_mB_j}\mright>\mright) \nonumber \\
     & + \dfrac{\mathrm{s}\mleft(4\alpha_i\mright)}{\mathrm{c}\mleft(4\alpha_i\mright)+\mathrm{c}\mleft(4\alpha_j\mright)} \mleft(\mleft<C_{\ell}^{B_iE_m}\mright>\mleft<C_{\ell}^{B_jB_n}\mright> + \mleft<C_{\ell}^{B_iB_n}\mright>\mleft<C_{\ell}^{E_mB_j}\mright>\mright) \nonumber\\
     & - \dfrac{\mathrm{s}\mleft(4\alpha_j\mright)}{\mathrm{c}\mleft(4\alpha_i\mright)+\mathrm{c}\mleft(4\alpha_j\mright)} \mleft(\mleft<C_{\ell}^{E_iE_m}\mright>\mleft<C_{\ell}^{E_jB_n}\mright> + \mleft<C_{\ell}^{E_iB_n}\mright>\mleft<C_{\ell}^{E_mE_j}\mright>\mright) \nonumber \\
     & + \dfrac{\mathrm{s}\mleft(4\alpha_j\mright)\mathrm{s}\mleft(4\alpha_n\mright)}{\mleft(\mathrm{c}\mleft(4\alpha_i\mright)+\mathrm{c}\mleft(4\alpha_j\mright)\mright)\mleft(\mathrm{c}\mleft(4\alpha_m\mright)+\mathrm{c}\mleft(4\alpha_n\mright)\mright)} \mleft(\mleft<C_{\ell}^{E_iE_m}\mright>\mleft<C_{\ell}^{E_jE_n}\mright> + \mleft<C_{\ell}^{E_iE_n}\mright>\mleft<C_{\ell}^{E_mE_j}\mright>\mright) \nonumber \\
     & + \dfrac{\mathrm{s}\mleft(4\alpha_i\mright)\mathrm{s}\mleft(4\alpha_m\mright)}{\mleft(\mathrm{c}\mleft(4\alpha_i\mright)+\mathrm{c}\mleft(4\alpha_j\mright)\mright)\mleft(\mathrm{c}\mleft(4\alpha_m\mright)+\mathrm{c}\mleft(4\alpha_n\mright)\mright)} \mleft(\mleft<C_{\ell}^{B_iB_m}\mright>\mleft<C_{\ell}^{B_jB_n}\mright> + \mleft<C_{\ell}^{B_iB_n}\mright>\mleft<C_{\ell}^{B_mB_j}\mright>\mright) \nonumber \\
     & - \dfrac{\mathrm{s}\mleft(4\alpha_j\mright)\mathrm{s}\mleft(4\alpha_m\mright)}{\mleft(\mathrm{c}\mleft(4\alpha_i\mright)+\mathrm{c}\mleft(4\alpha_j\mright)\mright)\mleft(\mathrm{c}\mleft(4\alpha_m\mright)+\mathrm{c}\mleft(4\alpha_n\mright)\mright)} \mleft(\mleft<C_{\ell}^{E_iB_m}\mright>\mleft<C_{\ell}^{E_jB_n}\mright> + \mleft<C_{\ell}^{E_iB_n}\mright>\mleft<C_{\ell}^{B_mE_j}\mright>\mright) \nonumber \\
     & \mleft. - \dfrac{\mathrm{s}\mleft(4\alpha_i\mright)\mathrm{s}\mleft(4\alpha_n\mright)}{\mleft(\mathrm{c}\mleft(4\alpha_i\mright)+\mathrm{c}\mleft(4\alpha_j\mright)\mright)\mleft(\mathrm{c}\mleft(4\alpha_m\mright)+\mathrm{c}\mleft(4\alpha_n\mright)\mright)} \mleft(\mleft<C_{\ell}^{B_iE_m}\mright>\mleft<C_{\ell}^{B_jE_n}\mright> + \mleft<C_{\ell}^{B_iE_n}\mright>\mleft<C_{\ell}^{E_mB_j}\mright>\mright)\mright] \nonumber
\end{align}

Due to the lack of a reliable model of the foregrounds cross-spectra, we require a good estimator of it from the observed power spectra. In section~\ref{subsec:covariance_matrix_model} we study several estimators and show that binning the observed power spectrum leads to competent results.

Let us assume: 
\begin{itemize}
    \item The rotation angles are small, i.e., $\alpha \ll 1 \rightarrow \mathrm{s}(\alpha) \sim \alpha$ and $\mathrm{c}(\alpha) \sim 1$.
    \item The covariance matrix $\mymatrix{C}$ does not depend on $\alpha$. To circumvent the mismatch induced by this approximation, we perform an iterative approach that updates the polarization angle in the covariance matrix with the one estimated in the previous step. 
\end{itemize} 
With these approximations, we achieve a linear system which enables us to obtain analytical equations from \eqref{eq:full_likelihood} to calculate the rotation angles. Moreover, the uncertainties can be evaluated from the Fisher matrix. This results in a very fast computational methodology. 

In order to maximize the likelihood, we obtain $N$ linear equations by taking the derivative of \eqref{eq:full_likelihood} with respect to each $\alpha$, and equate each of them to zero. Thus, the rotation angles are the solution of the following linear system:
\begin{equation}
    \mymatrix{\Omega}\myvector{\alpha} = \dfrac{1}{2}\myvector{\eta} \, , \label{eq:cross_angles_linear_system}
\end{equation}
where $\mymatrix{\Omega}$ is an $N\times N$ matrix and $\myvector{\eta}$ is an $N$ vector whose elements are:
\begin{align}
    \Omega_{ab} =& \sum\limits_{\ell=\ell_{m}}^{\ell_{M}}\mleft(\sum\limits_{i=1}^{N}\sum\limits_{m=1}^{N} C_{\ell}^{E_{i}E_{a}}\mymatrix{C}^{-1}_{\ell,iamb}C_{\ell}^{E_mE_{b}} - \sum\limits_{i=1}^{N}\sum\limits_{m=1}^{N} C_{\ell}^{E_{i}E_{a}}\mymatrix{C}^{-1}_{\ell,iabm}C_{\ell}^{B_{b}B_m}\mright) \nonumber \\
    & + \sum\limits_{\ell=\ell_{m}}^{\ell_{M}}\mleft(\sum\limits_{i=1}^{N}\sum\limits_{m=1}^{N} C_{\ell}^{B_{a}B_{i}}\mymatrix{C}^{-1}_{\ell,aibm}C_{\ell}^{B_{b}B_m} - \sum\limits_{i=1}^{N}\sum\limits_{m=1}^{N} C_{\ell}^{B_{a}B_i}\mymatrix{C}^{-1}_{\ell,aimb}C_{\ell}^{E_mE_{b}}\mright) \, ,
    \label{eq:omega_element_cross} \\
    \eta_{a} = & \sum\limits_{\ell=\ell_{m}}^{\ell_{M}}\mleft(\sum\limits_{i=1}^{N}\sum\limits_{m=1}^{N}\sum\limits_{n=1}^{N} C_{\ell}^{E_{i}E_{a}}\mymatrix{C}^{-1}_{\ell,iamn}C_{\ell}^{E_mB_n}  -\sum\limits_{i=1}^{N}\sum\limits_{m=1}^{N}\sum\limits_{n=1}^{N} C_{\ell}^{B_{a}B_i}\mymatrix{C}^{-1}_{\ell,aimn}C_{\ell}^{E_mB_n} \mright) \, .
    \label{eq:eta_element_cross}
  \end{align} 
The rotation angles' uncertainties are obtained from the Fisher matrix which is given as:
\begin{equation}
    \mymatrix{F} = 4\mymatrix{\Omega} 
    \label{eq:Fisher_cross}
\end{equation}

Although the formalism exposed uses the information from all auto- and cross-spectra among all channels, one can limit the information to only cross-spectra or auto-spectra (i.e., no correlation among different channels). In the former case, the angles can be estimated using equation~\ref{eq:cross_angles_linear_system} and equations~\ref{eq:omega_element_cross} and \ref{eq:eta_element_cross}, excluding the terms where  $i=a$, $m=b$ or $m=n$. In the latter case, with all the approximations made, the likelihood simplifies to 
\begin{equation}
    -2 \log \mathcal{L} \propto \sum\limits_{\ell=\ell_{m}}^{\ell_{M}} \mleft(\myvector{C}_{\ell,a}^{EB} - 2\myvector{\alpha}\mleft(\myvector{C}_{\ell,a}^{EE}-\myvector{C}_{\ell,a}^{BB}\mright)\mright)^T\mymatrix{M}^{-1}_{\ell}
    \mleft(\myvector{C}_{\ell,a}^{EB} - 2\myvector{\alpha}\mleft(\myvector{C}_{\ell,a}^{EE}-\myvector{C}_{\ell,a}^{BB}\mright)\mright)\, ,
    \label{eq:auto_likelihood}
\end{equation}
where $\myvector{C}^{XY}_{\ell,a}$ and $\myvector{\alpha}$ are $N$ vectors given by 
\begin{align}
    \myvector{C}^{XY}_{\ell,a} & = \mleft(C^{X_1Y_1}_{\ell}\,\,  C^{X_2Y_2}_{\ell} \,\, ... \,\, C^{X_NY_{N}}_{\ell} \mright)^{T} \, , \\
    \myvector{\alpha} & = \mleft(\alpha_1 \,\, \alpha_2 \,\, ... \,\, \alpha_N \mright)^{T} \, ,\\
\end{align}
and the covariance matrix $\mymatrix{M}_{\ell}$ is an $N\times N$ matrix whose elements are given as 
\begin{align}
    M_{\ell,im} = \dfrac{1}{2\ell +1} 
     & \mleft[\ensemble{C_{\ell}^{E_iE_m}}\ensemble{C_{\ell}^{B_iB_m}} + \ensemble{C_{\ell}^{E_iB_m}}\ensemble{C_{\ell}^{E_mB_i}}\mright. \nonumber \\  
     & + \tan(4\alpha_m)\mleft(\ensemble{C_{\ell}^{E_iB_m}}\ensemble{C_{\ell}^{B_iB_m}} - \ensemble{C_{\ell}^{E_mB_i}}\ensemble{C_{\ell}^{E_iE_m}}\mright) \nonumber \\
    & + \tan(4\alpha_i)\mleft(\ensemble{C_{\ell}^{E_mB_i}}\ensemble{C_{\ell}^{B_iB_m}} - \ensemble{C_{\ell}^{E_iB_m}}\ensemble{C_{\ell}^{E_iE_m}}\mright) \nonumber\\
    & + \dfrac{\tan(4\alpha_i)\tan(4\alpha_j)}{2}\mleft(\ensemble{C_{\ell}^{E_iE_m}}^2 + \ensemble{C_{\ell}^{B_iB_m}}^2\mright) \nonumber \\ 
    &\mleft . -\dfrac{\tan(4\alpha_i)\tan(4\alpha_j)}{2}\mleft(\ensemble{C_{\ell}^{E_iB_m}}^2 + \ensemble{C_{\ell}^{E_mB_i}}^2 \mright)\mright]  \, .
    \label{eq:auto_covariance_matrix}
\end{align}

Analogous to the full spectra case, the rotation angles are obtained after solving the following linear system:
\begin{equation}
    \mymatrix{\Theta}\myvector{\alpha} = \dfrac{1}{2}\myvector{\xi} \, , \label{eq:auto_angles_linear_system}
\end{equation}
where $\mymatrix{\Theta}$ is an $N\times N$ matrix and $\myvector{\xi}$ is an $N$ vector whose elements are:
\begin{align}
    \Theta_{ab} = \sum\limits_{\ell=\ell_{m}}^{\ell_{M}} \mleft(C_{\ell}^{EE_a}-C_{\ell}^{BB_a}\mright)\mymatrix{M}^{-1}_{\ell,ab}\mleft(C_{\ell}^{EE_b}-C_{\ell}^{BB_b}\mright)\, , 
    \label{eq:omega_auto_element} \\
    \xi_{a} = \sum\limits_{m=1}^{N}\sum\limits_{\ell=\ell_{m}}^{\ell_{M}} \mleft(C_{\ell}^{EE_a}-C_{\ell}^{BB_a}\mright)\mymatrix{M}^{-1}_{\ell,am}C_{\ell}^{EB_m} \, .
    \label{eq:eta_auto_element}
  \end{align} 
and the Fisher matrix is as
\begin{equation}
    \mymatrix{F}_a = 4\mymatrix{\Theta} .
    \label{eq:Fisher_auto}
\end{equation}

In section \ref{subsec:cross_cmbnoise} we study the differences among the results obtained when we use the total information available (cross and auto-spectra), only the cross-spectra information, or only the auto-spectra information.


\section{Simulations}
\label{sec:simulations}
We have generated 100 simulations of the observed sky with LiteBIRD (see table~\ref{tab:LiteBIRD_characteristics}). Each simulation has the following components: i) the sky signal, ii) the rotation angles and, iii) the instrumental noise. The sky signal is given as a collection of $N = 22$ ($\mymap{Q}$,$\mymap{U}$) pairs of frequency maps which contain the contribution of every significantly polarized physical emission. The rotation angles for a given simulation are generated from a distribution of zero mean and a covariance which correlates the angles within each instrument. Finally the instrumental noise is generated as white noise using  sensitivities similar to LiteBIRD's ($\sigma_{\nu}$ column of table~\ref{tab:LiteBIRD_characteristics}). For a given frequency $\nu$ the observed sky is calculated as follows:
\begin{equation}
    \begin{pmatrix}
		\mymap{Q}^{rot}_{\nu} \\
	    \mymap{U}^{rot}_{\nu}
	\end{pmatrix}
	=
	\begin{pmatrix}
		\cos(2\alpha_{\nu}) & -\sin(2\alpha_{\nu})\\
	    \sin(2\alpha_{\nu}) & \cos(2\alpha_{\nu})\\
	\end{pmatrix}
	\begin{pmatrix}
		\mymap{Q}_{\nu} \\
	    \mymap{U}_{\nu}
	\end{pmatrix}
	+ 
	\begin{pmatrix}
		\mymap{n}^{Q}_{\nu} \\
	    \mymap{n}^{U}_{\nu}
	\end{pmatrix}\, ,
	\label{eq:rotated_maps}
\end{equation}
where $\mymap{Q}_{\nu}$ ($\mymap{U}_{\nu}$) is the map with the $Q$ ($U$) sky signal, $\alpha_{\nu}$ is the rotation angle at the $\nu$ channel and, $\mymap{n}_{\nu}^{Q}$  and $\mymap{n}_{\nu}^{U}$ are white noise maps. In the following subsections, we explain how the sky signal and the rotation angles are simulated.

\begin{table}[tbp]
\centering
\begin{tabular}{|c|c|c|c|c|}
\hline
Instrument & $\nu$ (GHz) & FWHM (arcmin) & $\sigma_{\nu}$ ($\mu$K arcmin) \\
\hline
LFT & 40 & 70.5 &  37.43 \\
LFT & 50 & 58.5 & 33.46  \\
LFT & 60 & 51.1 &  21.32 \\
LFT & 68a & 41.6 & 19.91 \\
LFT & 68b & 47.1 & 31.76 \\
LFT & 78a & 36.9 & 15.56 \\
LFT & 78b & 43.8 & 19.14 \\
LFT & 89a & 33.0 & 12.28 \\
LFT & 89b & 41.5 & 28.77 \\
LFT & 100 & 30.2 & 10.34 \\
LFT & 119 & 26.3 & 7.69 \\
LFT & 140 & 23.7 & 7.24 \\
\hline
MFT & 100 & 37.8 & 8.48 \\
MFT & 119 & 33.6 & 5.70 \\
MFT & 140 & 30.8 & 6.39 \\
MFT & 166 & 28.9 & 5.56 \\
MFT & 195 & 28.0 & 7.04 \\
\hline
HFT & 195 & 28.6 & 10.50 \\
HFT & 235 & 24.7 & 10.80 \\
HFT & 280 & 22.5 & 13.80 \\
HFT & 337 & 20.9 & 21.95 \\
HFT & 402 & 17.9 & 47.44 \\
\hline
\end{tabular}
\caption{\label{tab:LiteBIRD_characteristics} LiteBIRD's channels specifications: the instrument where the channel is located, $\nu$ the channel frequency, FWHM of the channel beam, $\sigma_{\nu}$ the sensitivity \cite{SPIE_proceedings}. LFT, MFT and HFT refer to three LiteBIRD instruments and stand for Low, Medium and High Frequency Telescopes, respectively.}
\end{table}

\subsection{Sky Signal}
\label{subsec:simulations_sky_signal}

The multi-frequency sky signal maps are generated using the \texttt{HEALPix}\footnote{Hierarchical Equal Area isoLatitude Pixelization, \url{https://healpix.sourceforge.io/} \cite{gorski2005healpix}.} scheme at a resolution of $N_{side} = 512$ (LiteBIRD's expected pixelization). The sky signal is composed of the CMB signal as well as the principal polarized foregrounds. Each component is simulated as follows:

\paragraph{CMB.} CMB maps are drawn as Gaussian random realizations of theoretical angular power spectra. The power spectra are evaluated with the Boltzmann-solver \texttt{CAMB} \cite{lewis2011camb} using the cosmological parameters from Planck18 results \cite{aghanim2018planck} and a null tensor-to-scalar ratio.

\paragraph{Polarized foregrounds.} The primary polarized foreground sources are the synchrotron and the thermal dust emissions. We have generated three sets of multi-frequency maps of each foreground contaminant: 
\begin{itemize}
    \item Semi-realistic foregrounds (\texttt{s0d0}): the foregrounds are simulated using the \texttt{Python} software \texttt{PySM} \cite{thorne2017python}. In particular, we use the \texttt{s0} and \texttt{d0} models for the synchrotron and dust respectively. Both of these models assume constant spectral parameters over the sky. We have considered simple foreground models to avoid misidentifying residuals coming from improper foreground modelling as residuals from leftover rotation angles.
    \item Anisotropic Gaussian foregrounds (\texttt{AG}): we have generated anisotropic and Gaussian synchrotron and dust $\mymap{Q}$ and $\mymap{U}$ maps that satisfy the condition of having the same power spectra as the \texttt{s0d0} foregrounds\footnote{Note that these foregrounds do not need to follow the \texttt{s0d0} spectral laws.}. The procedure is the following:
    \begin{itemize}
        \item We calculate the $Q$ and $U$ local variance maps within $2^{\circ}$ radius disks of the synchrotron and dust emission at 40 and 337 GHz respectively. 
        \item Then, we generate synchrotron and dust template maps as Gaussian samples from the former maps. 
        \item To obtain the synchrotron and thermal dust maps at LiteBIRD's frequencies, we obtain the spherical harmonics ($e_{\ell m}$,$b_{\ell,m}$), and re-scale them using:
        \begin{align*}
            \hat{e}_{\ell m,\nu} & = \sqrt{\dfrac{E^{\texttt{s0d0}}_{\ell,\nu}}{E_{\ell}}} e_{\ell m}  \, ,\\
            \hat{b}_{\ell m,\nu} & = \sqrt{\dfrac{B^{\texttt{s0d0}}_{\ell,\nu}}{B_{\ell}}} b_{\ell m} 
        \end{align*}
        where $X^{\texttt{s0d0}}_{\ell,\nu}$ is either the \texttt{s0d0} synchrotron or thermal dust $X \in \{E,B\}$ power spectrum at the frequency $\nu$ and $X_{\ell}$ is the power spectrum of the template maps.
    \end{itemize} 
    \item Isotropic Gaussian foregrounds (\texttt{IG}): we generated isotropic and Gaussian synchrotron and dust $\mymap{Q}$ and $\mymap{U}$ maps as follows:
    \begin{itemize}
        \item We generated two $Q$ and two $U$ template maps by drawing Gaussian samples from the standard normal distribution.
        \item Then, the $Q$ and $U$ synchrotron and thermal dust maps at LiteBIRD's frequencies are obtained by re-scaling these template maps by the standard deviation of the corresponding \texttt{s0d0} foreground map.
    \end{itemize}
\end{itemize} 
To test the assumption of having a null $EB$ foregrounds contribution, we have generated analogous versions of the \texttt{AG} and \texttt{IG} foregrounds with zero $EB$ by removing their $EB$ contribution. These foregrounds are denoted by \texttt{AGnEB} and \texttt{IGnEB}, respectively. For this purpose, we modified the spherical harmonics coefficients of the $B$-mode ($b_{\ell m}$) as follows:
    \begin{equation}
    \hat{b}_{\ell m}  = \beta_{\ell} b_{\ell m} + \gamma_{\ell} e_{\ell m} \, ,\\
    \end{equation}
    where $e_{\ell m}$ are the spherical harmonics coefficients of the $E$-mode and:
    \begin{align}
        \beta_{\ell} & = +\sqrt{\dfrac{C_{\ell}^{BB}C_{\ell}^{EE}}{C_{\ell}^{BB}C_{\ell}^{EE}-\mleft(C_{\ell}^{EB}\mright)^2}}\, ,\\
        \gamma_{\ell} & = -\dfrac{C_{\ell}^{EB}}{C_{\ell}^{EE}}\beta_{\ell}\, .
    \end{align}
These conditions satisfy:
    \begin{align}
        \mleft<\hat{b}_{\ell m},e_{\ell m}\mright> & = C_{\ell}^{E\hat{B}} = \beta_{\ell}C_{\ell}^{EB} + \gamma_{\ell}C_{\ell}^{EE} = 0 \, , \\
        \mleft<\hat{b}_{\ell m},\hat{b}_{\ell m}\mright> & = C_{\ell}^{\hat{B}\hat{B}} = \beta_{\ell}^2C_{\ell}^{BB} + \gamma_{\ell}^2C_{\ell}^{EE} + 2\beta_{\ell}\gamma_{\ell}C_{\ell}^{EB} = C_{\ell}^{BB} \, .
    \end{align}
Finally, both the CMB and foregrounds maps are convolved with Gaussian beams using the corresponding FWHM specified in table~\ref{tab:LiteBIRD_characteristics}. 

\subsection{Rotation Angles}
\label{subsec:simulations_rotation_angles}


In this study, we have assumed that the rotation angles within a given telescope (e.g., the LFT) are correlated, but they are uncorrelated among different telescopes. In \cite{requirements} they obtained the maximum uncertainty on the polarization angle calibration that does not introduce a bias in the $r$ determination, taking into account possible correlations among different frequency channels, i.e., in the scenario we have assumed. We use values ($\sigma_{\alpha,\nu}$) compatible to those from  Case 0 reported in \cite{requirements} to obtain the distribution from which we sample the rotation angles.

The rotation angles distribution is a multi-variate Gaussian distribution with zero mean and the following covariance matrix:
\begin{equation}
    \mymatrix{C}_{\alpha} = 
    \begin{pmatrix}
    M_{LFT} & 0 & 0 \\
    0 & M_{MFT} & 0 \\
    0 & 0 & M_{HFT} \\
    \end{pmatrix}
    \, ,
    \label{eq:channels_rotation_angles_correlation_matrix}
\end{equation}
where $M_{ins}$ is the instrument covariance matrix whose elements are $M_{ins,ij} = \rho_{ij} \sigma_{\alpha,i}\sigma_{\alpha,j}$, $\sigma_{\alpha,i}$ is the miscalibration angle uncertainty of the $i$-th channel, and $\rho_{ij}$ is the correlation coefficient among the channels $i$ and $j$. Since we assumed full correlations among the channels at a given instrument $\rho_{ij} = 1$ $\forall i,j$. Note that the most stringent requirements are found when the channels are fully correlated. Therefore, the uncertainties are lower than in other cases leading to smaller rotation angles. However, it is worth mentioning that the requirements do not provide the actual calibration uncertainties. Moreover, the expected calibration uncertainties are smaller than the requirements for the least favourable scenario (i.e., no correlation among channels\cite{requirements}).   

One can generate samples with the covariance properties of $\mymatrix{C}_{\alpha}$ using  the  Cholesky  decomposition. $\mymatrix{C}_{\alpha}$ is a real positive-definite symmetric matrix, hence the lower-triangular matrix $\mymatrix{L}$ that satisfies $\mymatrix{C}_{\alpha} = \mymatrix{L}\mymatrix{L}^{T}$ can be obtained by applying the Cholesky decomposition. Then, if $\myvector{x}$ is an $N$ vector whose components are  independent random samples derived from the standard Gaussian distribution, $\mymatrix{L}\myvector{x}$ is a sample vector from the $\mathcal{N}(0,\mymatrix{C}_{\alpha})$ distribution. Thus, the rotation angles for a given simulation are obtained as 
\begin{equation}
    \myvector{\alpha} = \mymatrix{L}\myvector{x} \, .
\end{equation}

\section{Methodology Performance}
\label{sec:performance}


In this section, we study the performance of the methodology introduced in section~\ref{sec:methodology} using the 100 multi-frequency simulations described in section~\ref{sec:simulations}. Unless stated otherwise, we use the LiteBIRD-like simulations i.e., CMB+\texttt{s0d0}+Noise.  In particular, we discuss the difficulties that arise from computing the covariance matrix from the observed spectra, and how to alleviate this problem, in section~\ref{subsec:covariance_matrix_model}, as well as checking the convergence of the algorithm and the compatibility of the Fisher uncertainties with the error bars that can be estimated from the dispersion of the simulations in section~\ref{subsec:convergence_uncertainty}. As a consistency test, in section~\ref{subsec:total_vs_individual_telescopes} we also compare the results that can be obtained from a telescope-by-telescope analysis with the ones coming from the full set of channels. Those studies are performed using only the information from the auto-spectra. Finally, in section~\ref{subsec:cross_cmbnoise}, we compare the results recovered in the three information cases considered: i) when both the cross- and auto-spectra information is used, ii) when only the cross-spectra are taken into account and, iii) when only auto-spectra are considered.

\subsection{Building the Covariance Matrix from Observed Spectra} 
\label{subsec:covariance_matrix_model}

As previously mentioned, in the absence of a reliable model of the angular power spectra of Galactic foregrounds, we must calculate the covariance matrix from the observed data. Although this might be seen as a positive outcome since it makes our methodology model-independent, in practice, building the covariance matrix from observed spectra can introduce significant uncertainty and numerical instability. We have explored several approaches to mitigate this problem, namely, adjusting the multipole range to avoid the contribution of noise-dominated angular scales as well as using different binning strategies.

The power of our methodology lies on the combination of information from various frequency channels. However, each of these frequency bands has associated a different instrumental beam and noise, and thus, although the smaller angular scales might be accessible for some channels, they will be completely noise-dominated in others. The inclusion of these noise-dominated scales will severely affect the calculation of the covariance matrix, consequently limiting the performance of the whole methodology. Therefore, the first action to take is to tune the multipole range to balance the contribution of each channel. Note that, although the discussion presented below is specific to LiteBIRD, the same optimization process can be applied to any other instrumental configuration.

\begin{figure}
\centering 
\includegraphics[width=1.0\textwidth]{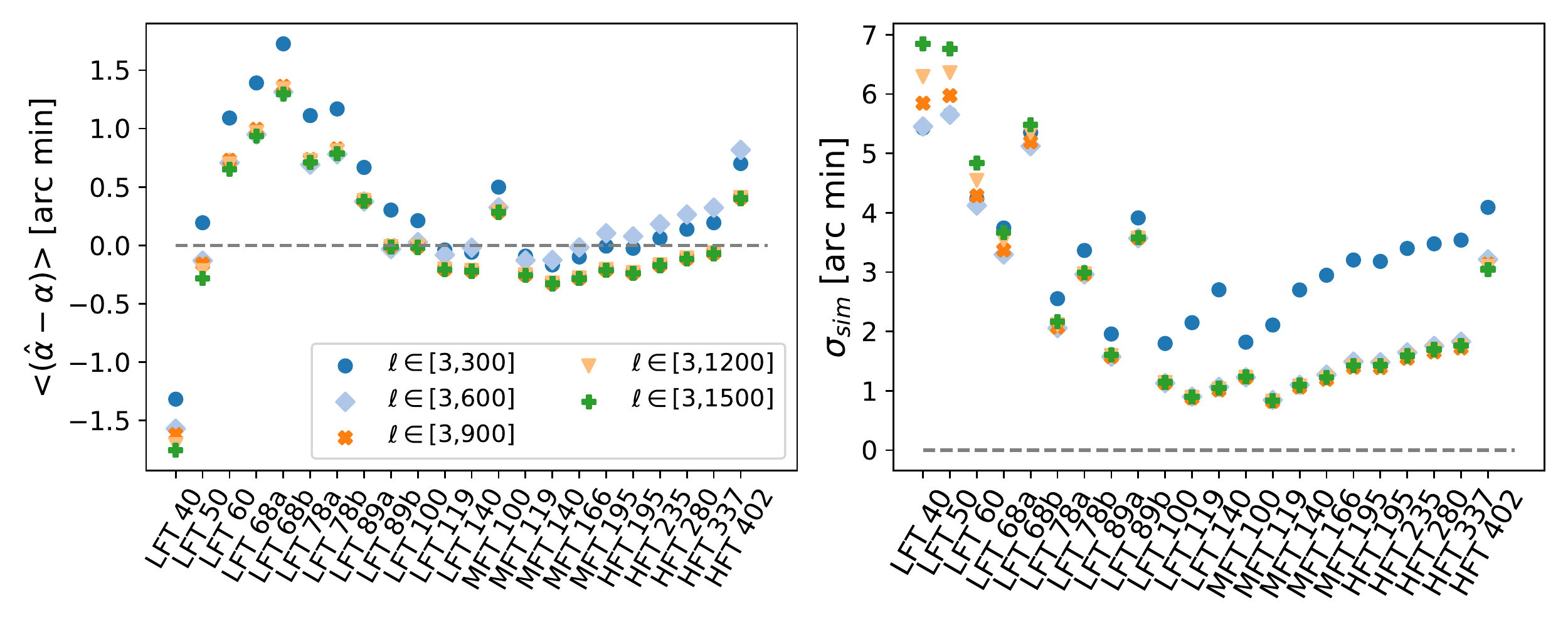}
\caption{\label{fig:ell_range_comparison_1} Left panel: Mean difference between the estimated and the genuine rotation angle per channel for different $\ell_M$. Right panel: $\sigma_{\mathrm{sim}}$ uncertainty obtained as the simulations dispersion.} 
\end{figure}

\begin{figure}
\centering 
\includegraphics[width=1.0\textwidth]{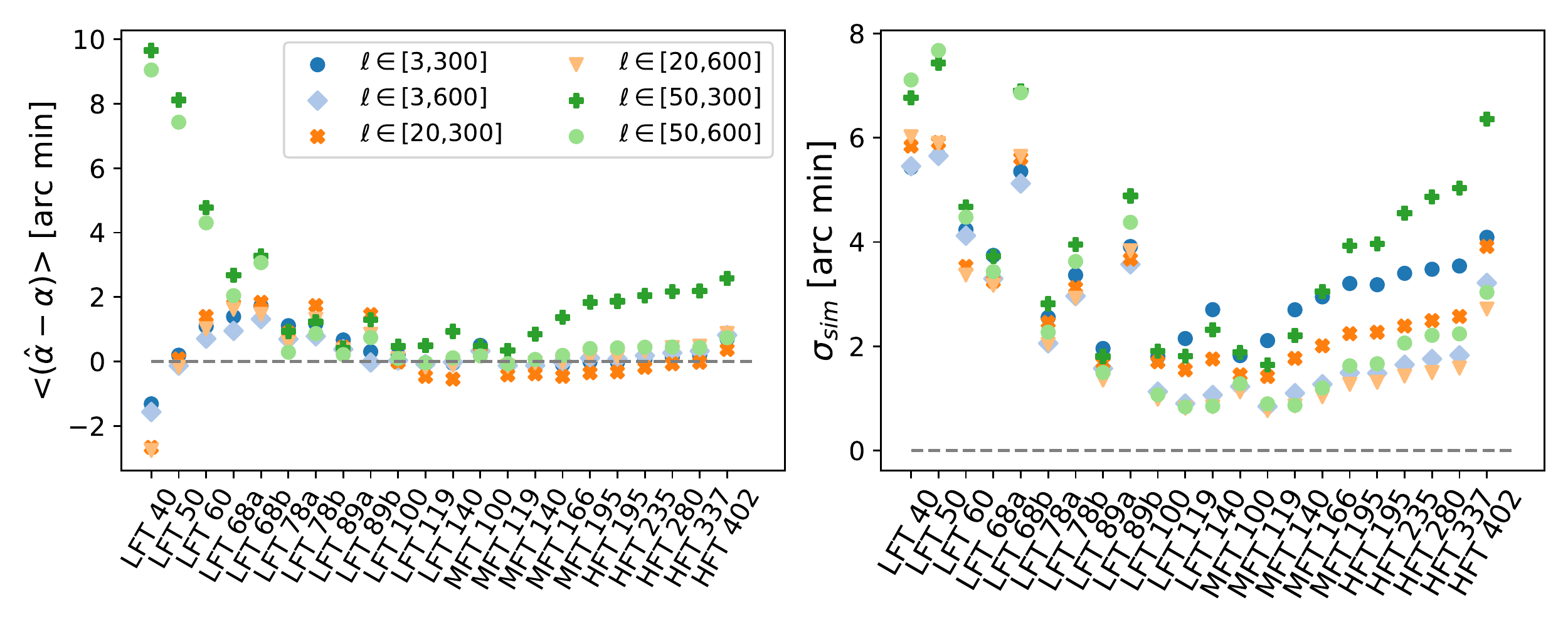}
\caption{\label{fig:ell_range_comparison_2} Left panel: Mean difference between the estimated and the genuine rotation angle per channel for different $\ell_m$ and $\ell_{M}$ cuts. Right panel: $\sigma_{\mathrm{sim}}$ uncertainty obtained as the simulations dispersion.}
\end{figure}

Looking at the current configuration of LiteBIRD's frequency bands (see table~\ref{tab:LiteBIRD_characteristics}), the FWHM of its beams varies between $70$ to $18$ arcmin, with a typical resolution across channels of around $30$ arcmin. Angular scales below the beam size are severely suppressed. Therefore, although the multipole range allowed by the expected LiteBIRD map resolution is approximately [3,1500]\footnote{We decided to omit $\ell=2$ because it was problematic for the inversion of the covariance matrix}, we have explored limiting the maximum multipole to $\ell_{M}=\{300, 600, 900, 1200, 1500\}$. For completeness, we also evaluated the impact of applying $\ell_{m}=\{20, 50\}$ cuts to the performance of the estimation. Figures \ref{fig:ell_range_comparison_1} and \ref{fig:ell_range_comparison_2} show the effect that different combinations of these $\ell_{m}$ and $\ell_{M}$ intervals have on the bias per channel of the rotation angles estimates and on the $\sigma_{\mathrm{sim}}$ uncertainty calculated as the simulations' dispersion.

As can be seen in figure \ref{fig:ell_range_comparison_1}, exploiting the full multipole range allowed by the beam resolution is crucial to obtain a precise determination of rotation angles, but including information from scales beyond the beam resolution does not necessarily grant any additional signal-to-noise. In this way, fixing $\ell_{M}$ to match the angular resolution of the widest beam ($\ell_{M}=600$) improves the performance on high-resolution channels without significantly deteriorating the results obtained for the low frequency bands of the LFT. In figure \ref{fig:ell_range_comparison_2}, we observe that cutting $\ell_{m}$ severely reduces the accuracy at the low frequency channels of the LFT. This is expected since, due to their lower resolution, most of the information from the LFT channels comes from the larger scales. On the contrary, the uncertainty on the rotation angle estimates from MFT bands is almost insensitive to $\ell_{m}$ cuts because, thanks to their narrower beams and better sensitivities, the main contribution comes from $\ell\gtrsim150$ scales. Note that, because of their higher instrumental noises and broader beams, the lower frequencies of the LFT channels are expected to yield worse results. HFT channels behave as those from the MFT as they have narrow beams and share similar noise levels, except the 402 GHz channel where the sensitivity is worse, leading to larger uncertainties and less accurate estimates. Therefore, we decided to choose the multipole range that optimizes the rotation angle estimations at cosmological channels. According to these results, the best option would then be to work within the $\ell\in[3,600]$ multipole interval. Hereinafter, the results will be obtained using this multipole interval. 

\begin{figure}
\centering 
\includegraphics[width=1.0\textwidth]{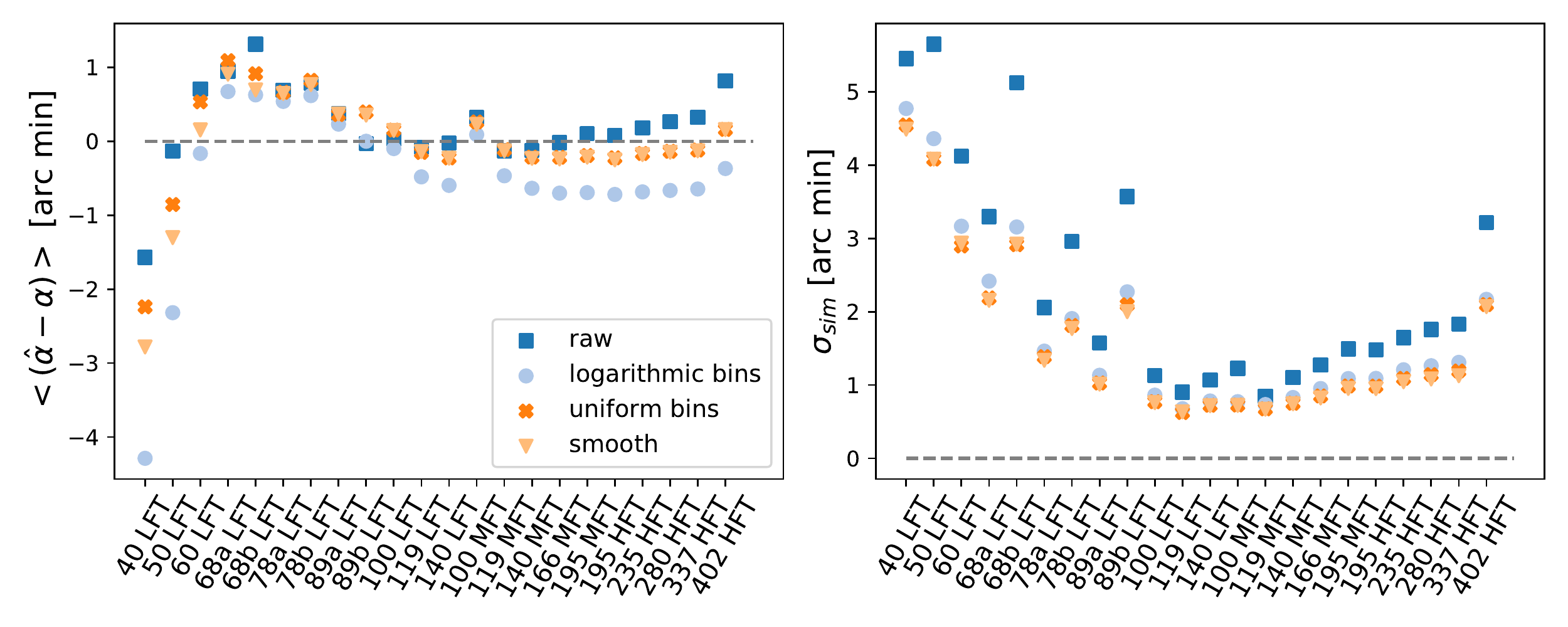}
\caption{\label{fig:smooth_binning} Left panel: Mean difference between the estimated and the genuine rotation angle per channel obtained using the raw power spectra (dark blue squares), smoothed power spectra (light orange triangles), logarithmically binned power spectra (light blue circles) and uniformly binned power spectra (orange crosses). Right panel: $\sigma_{\mathrm{sim}}$ uncertainty obtained as the simulations dispersion. The multipole range considered is $\ell \in [3,600]$. }
\end{figure}

So far, we have worked with the raw observed angular power spectra. In order to further reduce the noise in the calculation of the covariance matrix, and achieve an overall improvement of the results, we have also tried to smooth or bin the observed angular power spectra. To smooth the spectra, we have convolved the power spectra twice with two different square window functions, the first window function of $5\ell$ length, and the second of $10\ell$ length. We have also studied different binning schemes, e.g., using either uniform or logarithmically distributed bins (larger bins towards smaller angular scales).  In figure~\ref{fig:smooth_binning} we show the optimal configuration for the uniform and logarithmic bins that we found (50 and 10 bins in the uniform and logarithm case respectively), as well as the results obtained by smoothing the spectra. 

From figure~\ref{fig:smooth_binning} it is clear that either binning or smoothing the observed angular power spectra improves the uncertainty of the rotation angle estimates across all frequency bands. Although this improvement in precision does not necessarily translate into an improvement in accuracy (in fact, the mean errors obtained when applying a logarithmic binning tend to be greater than those obtained without applying any binning or smoothing), smoothing and uniformly binning the observed spectra does indeed improve the overall performance of the methodology since, for cosmological channels, we obtain a better precision and comparable, or even slightly better, mean errors. Albeit smoothing the observed spectra seems to be a good option according to these results, it leads to larger discrepancies between Fisher uncertainties and the error bars that can be estimated from the simulations dispersion, aspects that will be discussed in the next subsection. Thus we favor uniformly binning the spectra over smoothing it. In the following all the results are obtained using the uniform binning scheme except as otherwise indicated.

\subsection{Convergence and Uncertainty Estimation}
\label{subsec:convergence_uncertainty}

\begin{figure}
\centering 
\includegraphics[width=\textwidth]{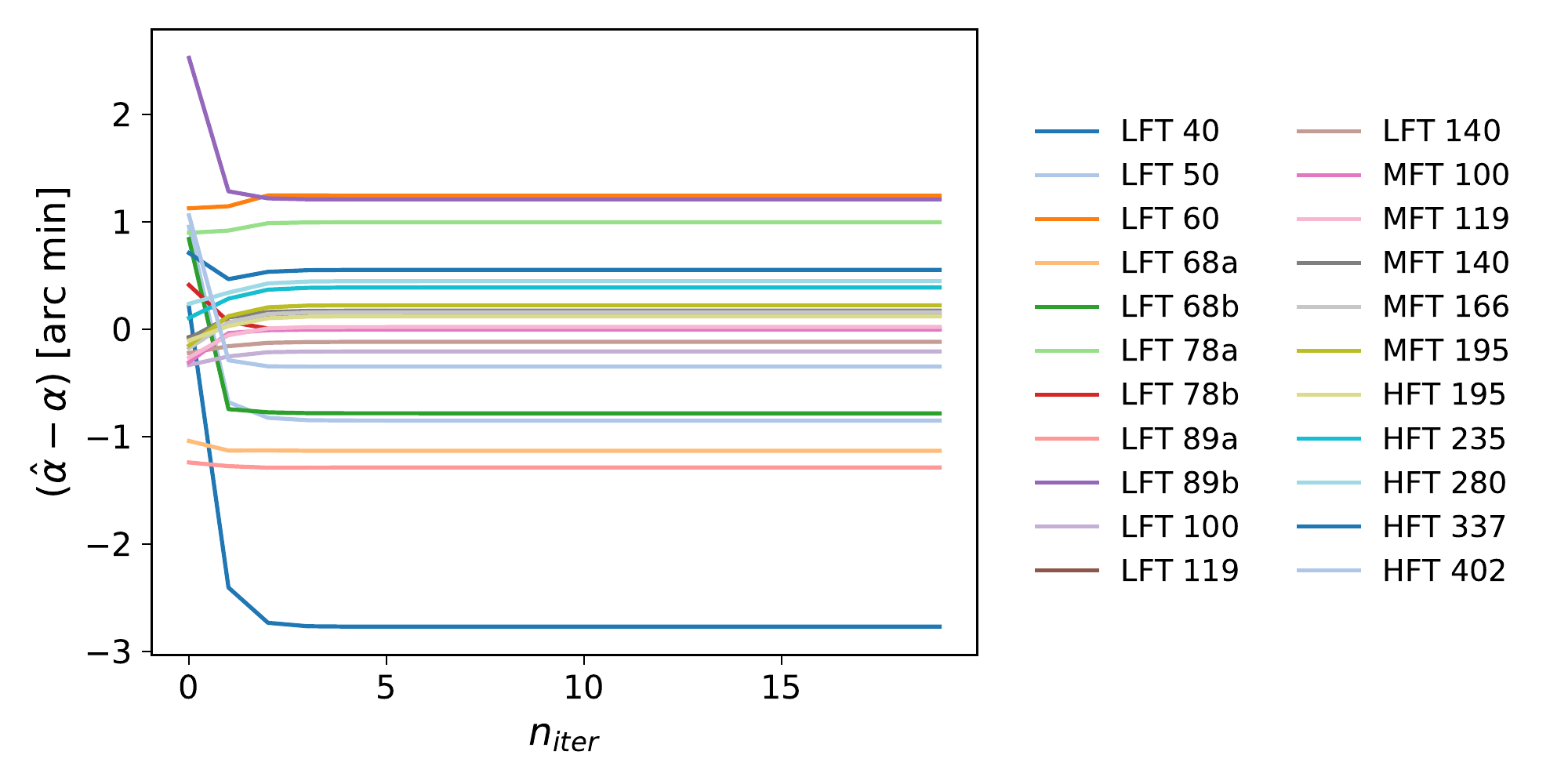}
\caption{\label{fig:iterations} 
Difference between the estimated and the true rotation angle per iteration for one simulation using the default configuration (50 uniformly distributed bins within the [3,600] multipole range). The results are shown for all LiteBIRD channels. 
}
\end{figure}

One of the advantages that our method presents with respect to other similar methodologies \cite{minami2019simultaneous,yuto_partial-sky,yuto_cross-spectra} based on the same principles, but relying on MCMC implementations,  is its speed. Although we have designed an iterative algorithm, once the input frequency maps are loaded and their spectra are calculated, each iteration only demands the inversion of the updated covariance matrix and  solving the linear system. The inversion of the covariance matrix can be speeded up since the inversion of a $NN_\ell\times NN_\ell$\footnote{Notice that we are using auto-spectra, in the cross (total) spectra the covariance matrix is an $N(N-1)N_\ell\times N(N-1)N_\ell$ ($N^2N_\ell\times N^2N_\ell$) matrix.} matrix of diagonal $N_\ell\times N_\ell$ boxes is equivalent to the inversion of $N_\ell$ matrices of $N\times N$. In addition, as shown in figure \ref{fig:iterations}, for the chosen binning configuration (50 uniformly distributed bins within $\ell\in[3,600]$), the algorithm quickly converges to its final estimation after a couple of iterations. For some of the binning configurations and multipole intervals that lead to poorer angle estimates discussed in section \ref{subsec:covariance_matrix_model}, a larger number of iterations (of the order of tens) would be needed to converge to a stable solution. Therefore, the only time-consuming step is the initial reading and processing of frequency maps. This increase in speed does not entail a loss of accuracy since, as demonstrated in \cite{RAC}, both types of methodologies achieve comparable results.

As mentioned in section~\ref{sec:methodology}, our method can provide an estimate of the uncertainties associated to the best-fit values using the Fisher matrix approximation. Alternatively, the uncertainty can be characterized by applying the method to a set of simulations, and calculating the dispersion of the errors in the estimation of the rotation angles. If the likelihood is well behaved and posterior distributions are Gaussian-like, the $\sigma$ obtained from the simulations dispersion and from the Fisher analysis should match. Checking the compatibility of these two estimates of the uncertainty is another of the tests we carried to evaluate the methodology performance.

\begin{figure}
    \hspace*{-.015\textwidth}
    \begin{minipage}{.5\linewidth}
        \centering 
        \includegraphics[width=\textwidth]{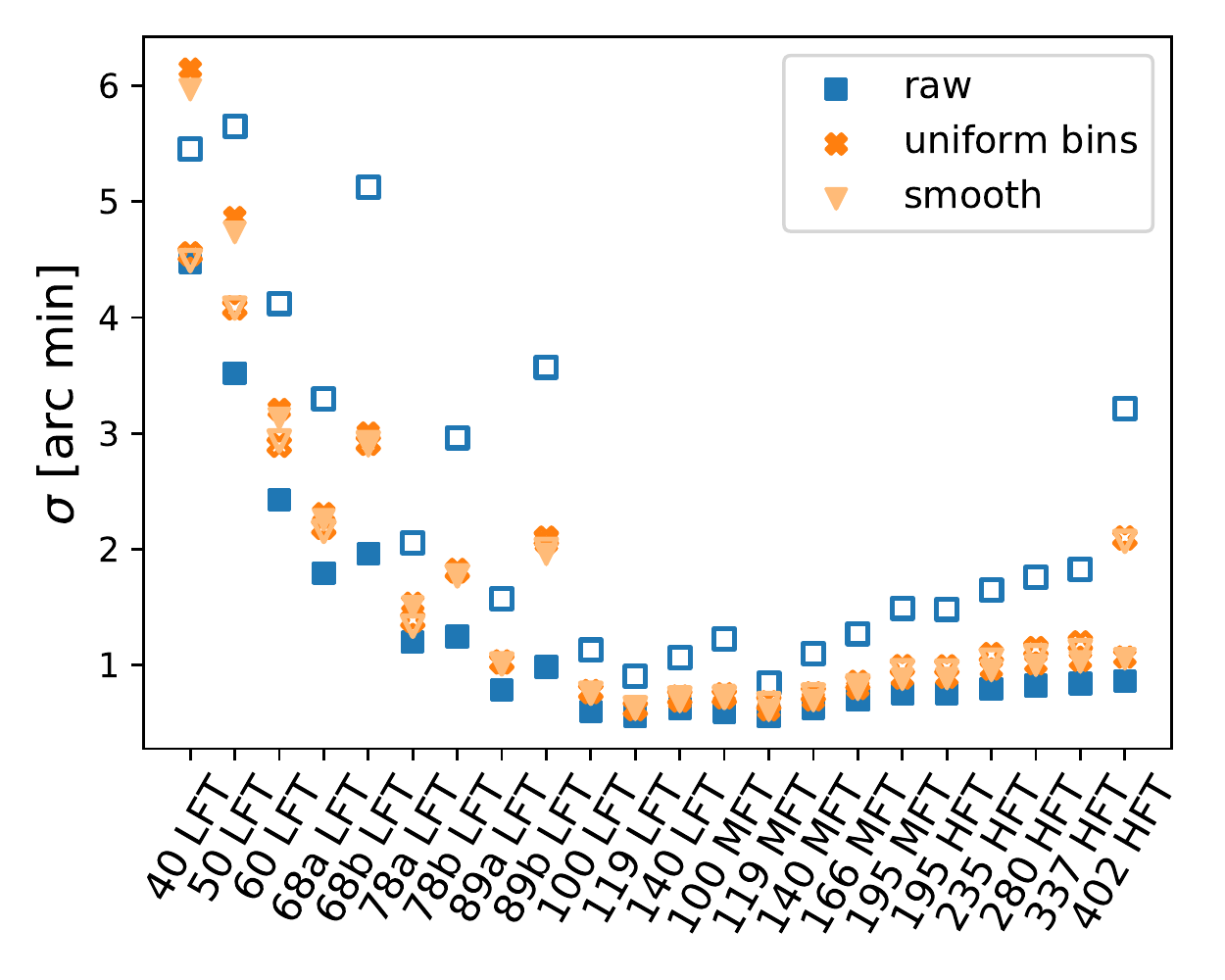}
        \caption{\label{fig:simDips_vs_fisher} Comparison between the Fisher uncertainties (solid markers) and simulations uncertainties (empty markers) for some of the binning configurations and for the optimal multipole interval discussed in section \ref{subsec:covariance_matrix_model}.}
    \end{minipage}
    \hspace{.03\textwidth}
    \begin{minipage}{.5\linewidth}
        \centering 
        \includegraphics[width=\textwidth]{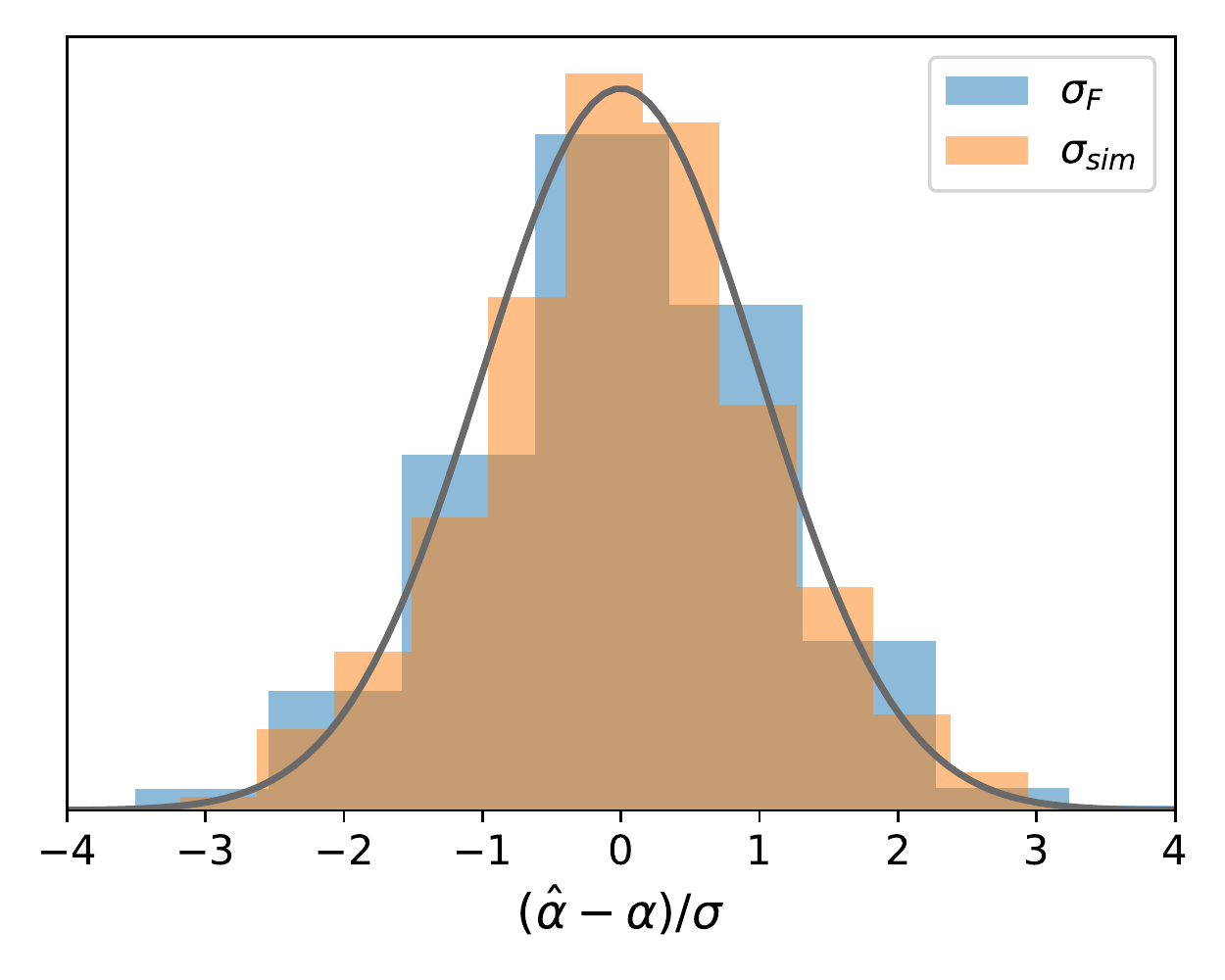}
        \caption{\label{fig:distribution} Distribution of the estimated and true rotation angles difference normalized by either the Fisher or the simulations uncertainty. The solid line shows the density function of the standard normal distribution.}
\end{minipage}
\end{figure}


In figure \ref{fig:simDips_vs_fisher} we compare the $\sigma$ obtained from the simulations dispersion and from a Fisher analysis for some of the binning and smoothing configurations discussed in section \ref{subsec:covariance_matrix_model}. As can be seen in this figure, although binning and smoothing slightly increases the Fisher uncertainties, it also helps to make both estimations more compatible. Given that uniformly binning the spectra, in addition to reducing the bias and the error bars estimated from simulations analysis (see section \ref{subsec:covariance_matrix_model}), also seems to yield the best agreement between Fisher uncertainties and the simulations dispersion, we chose this approach as our default configuration. Because of this good agreement we can trust that, when applied to different data, the Fisher matrix will give us a good estimate of the uncertainties associated to the best-fit values. We observe that, when  the power spectra are either binned or smoothed, the simulations uncertainty is smaller than Fisher's at low frequencies. These results seem to contradict the Cramer-Rao Bound. However, the Cramer-Rao Bound only applies when the likelihood is correctly defined. In our case, we do not have a model of the true covariance matrix, and these results show that the approximations made in its calculations are not sufficient. In section~\ref{subsec:cross_cmbnoise} we analyze more in depth the role that the covariance matrix plays in the rotation angles estimations.

\begin{figure}
\centering
\includegraphics[angle=270,width=.95\textwidth]{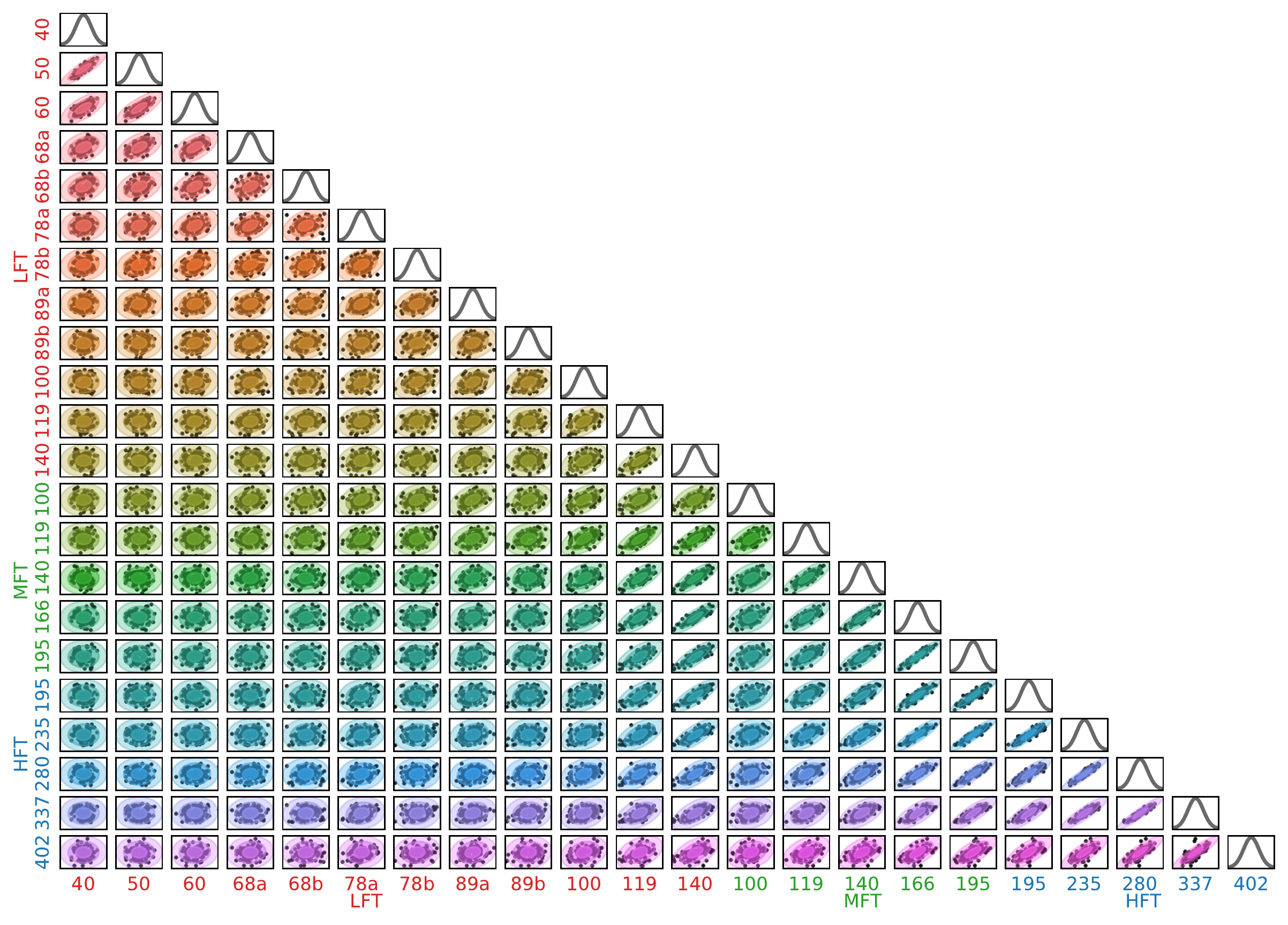}
\caption{\label{fig:triangular_plot_telescopes} Correlation matrix of the rotation angles. On top of the point cloud from the simulations, we show the $1\sigma$, $2\sigma$ and $3\sigma$ confidence contours obtained from the Fisher analysis.} 
\end{figure}

Figure \ref{fig:distribution} shows the distribution of the difference between the estimated and the genuine rotation angles normalized by either the Fisher or simulations uncertainty when the uniform binning scheme is used. All channels and simulations are included in the distribution. The probability density function of the standard normal distribution is also shown for comparison. Both distributions have a mean and standard deviation compatible with the standard normal distribution showing that our results are unbiased and that Fisher provides a good estimation of the error. Furthermore, as can be seen in figure \ref{fig:triangular_plot_telescopes}, Fisher confidence contours also reproduce correctly the correlations existing between different frequency bands (off-diagonal terms of the covariance matrix).  As can be seen in this figure, in general, angle estimates are fairly independent across frequency channels, showing only a slight correlation between adjacent frequency bands, with the exception of  dust-dominated bands ($\geq 100$GHz channels of the LFT, MFT and HFT) where angles are strongly correlated amongst themselves.

\subsection{Telescope-by-telescope Analysis}
\label{subsec:total_vs_individual_telescopes}

\begin{figure}
\centering 
\includegraphics[width=1.0\textwidth]{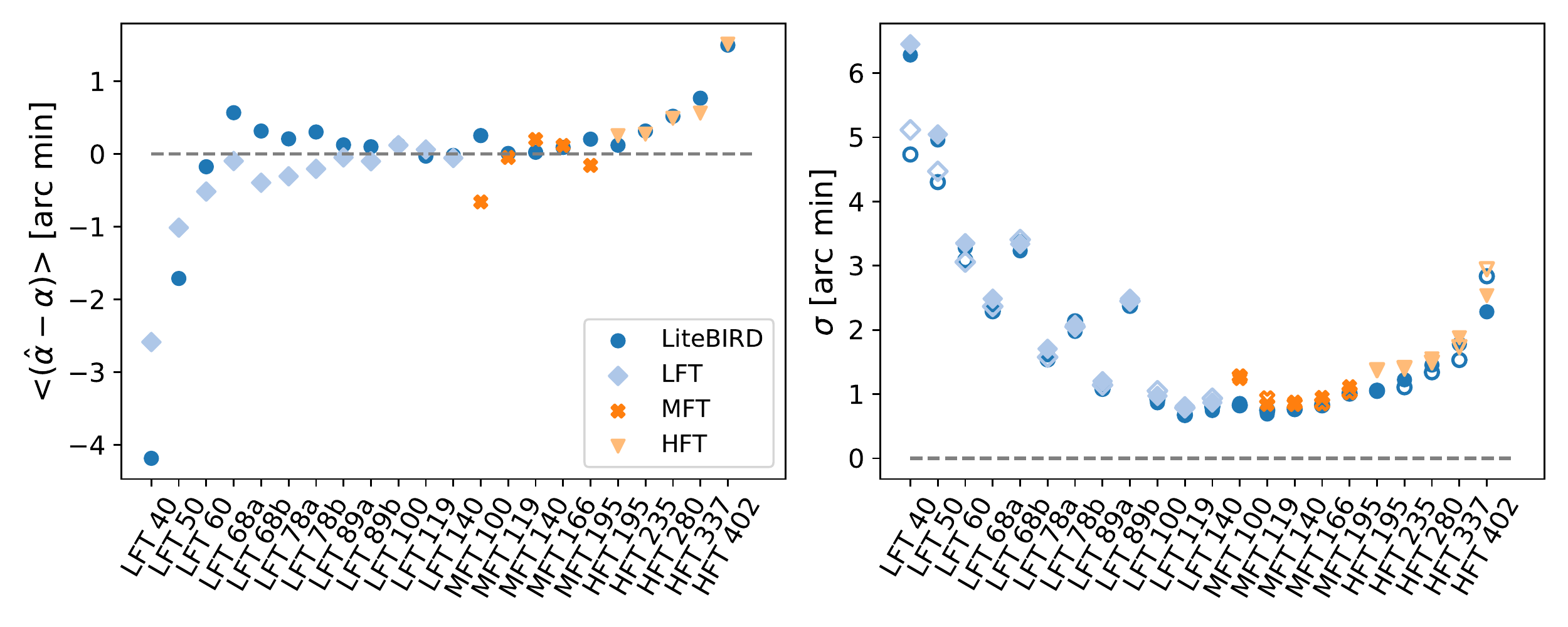}
\caption{\label{fig:telescope_vs_all} Left panel: Mean difference between the estimated and the genuine rotation angle per channel obtained from a telescope-by-telescope analysis  compared to the previous results obtained from the joint analysis of all frequency bands. Right panel: Fisher (solid markers) and simulations (empty markers) uncertainties. }
\end{figure}
Until now we have estimated the rotation angles of all LiteBIRD's frequency channels at once. However, a telescope-by-telescope determination of rotation angles will also be a reasonable approach since detectors are physically distributed in distinct telescopes, which might be systematically rotated in different orientations, and they target very different frequency ranges, each dealing with different weights of the Galactic foregrounds. Therefore, checking the compatibility of the results obtained from a telescope-by-telescope analysis with those from the joint analysis of all frequency bands will be, both a good consistency test, and a way to identify which telescope benefits more from the extra information coming from the others.

Figure \ref{fig:telescope_vs_all} shows the comparison of the mean error and uncertainties in the estimation of rotation angles obtained from a telescope-by-telescope analysis with the previous results obtained from the joint analysis of all frequency bands (50 uniformly distributed bins between $\ell\in[3,600]$). Like it was done in section \ref{subsec:covariance_matrix_model} for the full set of channels, the multipole range and binning strategy has been tailored to optimize the performance of each individual telescope (LFT: $\ell \in [3,600]$, MFT: $\ell \in [3,1200]$ and, HFT: $\ell \in [20,900]$ and uniform binning). This figure shows that the uncertainty in the estimation of rotation angles is slightly higher in the telescope-by-telescope analysis compared to the joint analysis. Regarding the bias, in the MFT and HFT the results are very similar but a small improvement is observed in the LFT channels. In light of these results, we conclude that a telescope-by-telescope analysis could constitute a good consistency test to check possible inconsistencies within a specific telescope.

\subsection{Comparison among Different Likelihood Estimators}
\label{subsec:cross_cmbnoise}

\begin{figure}
    \centering
    \includegraphics[width=\linewidth]{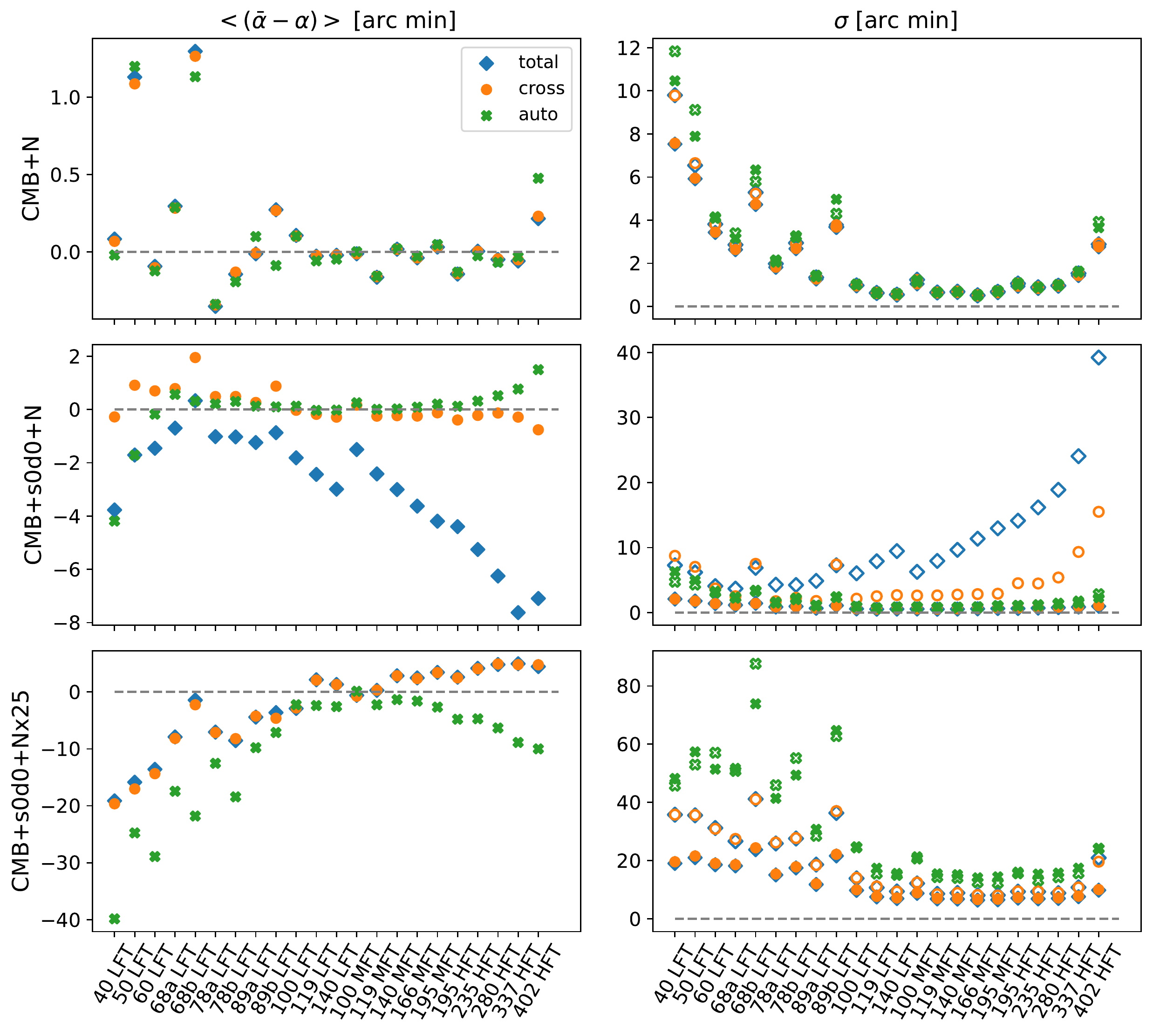}
    \caption{Estimations and uncertainties recovered when all spectra (total), only cross-spectra (cross), and only auto-spectra (auto) are considered in the CMB+N, CMB+\texttt{s0d0}+N and, CMB+\texttt{s0d0}+N$\times$25 sky scenarios. Left panels: mean difference between the angle estimates and the genuine angles. Right panels: Fisher uncertainties (solid markers) and simulations uncertainties (empty markers).}
    \label{fig:comparison_total_cross_auto}
\end{figure}

\begin{figure}
    \centering
    \includegraphics[width=\linewidth]{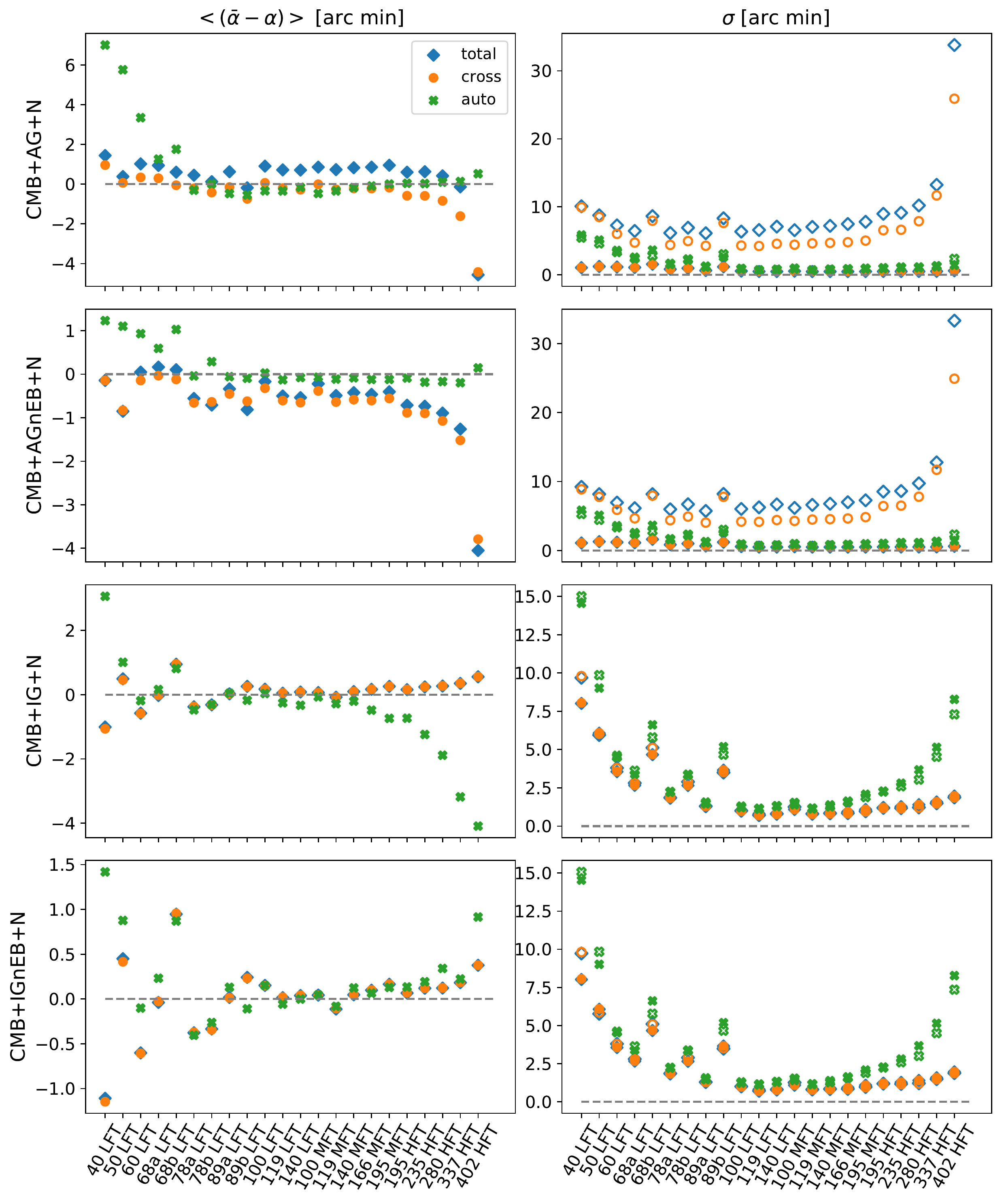}
    \caption{Estimations and uncertainties recovered when all spectra (total), only cross-spectra (cross), and only auto-spectra (auto) are considered in the CMB+\texttt{AG}+N, CMB+\texttt{AGnEB}+N, CMB+\texttt{IG}+N and, CMB+\texttt{IGnEB}+N sky scenarios. Left panels: mean difference between the angle estimates and the genuine angles. Right panels: Fisher uncertainties (solid markers) and simulations uncertainties (empty markers).}
    \label{fig:comparison_total_cross_auto_sim}
\end{figure}

In section~\ref{sec:methodology} we presented the formalism for three different estimators:
i) when both the cross-spectra and auto-spectra  information is used (total), ii) when only the cross-spectra is examined (cross) and, iii) when only the auto-spectra are considered (auto). Here, we have studied the accuracy of the rotation angles estimates and their uncertainties obtained in each case for the following sky models:
\begin{itemize}
    \item CMB+N: 100 simulations of the sky with only CMB and white noise.
    \item CMB+\texttt{s0d0}+N: 100 simulations of the sky with CMB, the \texttt{s0d0} foregrounds and white noise.
    \item CMB+\texttt{s0d0}+N$\times$25: 100 simulations of the sky with CMB, the \texttt{s0d0} foregrounds and white noise. The noise dispersion is scaled by a factor of 25.
    \item CMB+\texttt{AG}+N: 100 simulations of the sky with CMB, an anisotropic Gaussian foregrounds realization, and white noise.
    \item CMB+\texttt{AGnEB}+N: 100 simulations of the sky with CMB, the previous anisotropic Gaussian foregrounds realization without its $EB$ angular power spectrum contribution, and white noise.
    \item CMB+\texttt{IG}+N: 100 simulations of the sky with CMB, an isotropic Gaussian foregrounds realization, and white noise. 
    \item CMB+\texttt{IGnEB}+N: 100 simulations of the sky with CMB, the previous isotropic Gaussian foregrounds realization without its $EB$ angular power spectrum contribution, and white noise. 
\end{itemize}
The results are shown in figures~\ref{fig:comparison_total_cross_auto} and \ref{fig:comparison_total_cross_auto_sim}.

In the ideal scenario where there are no foregrounds (CMB+N, first row of figure~\ref{fig:comparison_total_cross_auto}) we find that the total, cross, and auto results show similar accuracies regarding the rotation angles estimates. Besides, the Fisher and simulations uncertainties agree in general quite well for the different cases. The uncertainties obtained in the cross and total cases are slightly lower at the lowest and highest frequencies with respect to the auto case. This is simply a consequence of having $N$ times ($N-1$ times) more information in the total (cross) case than in the auto case. 

When \texttt{s0d0} foregrounds are included (second row of figure~\ref{fig:comparison_total_cross_auto}) we observe that the estimates when the total information is used are biased. This effect arises due to numerical errors introduced by an improper characterization of the foregrounds in the covariance matrix, further  enhanced after inverting the covariance matrix. Those numerical errors are significantly mitigated when either only the cross or the auto spectra are considered. However, the Fisher uncertainties obtained in the cross case underestimate considerably the uncertainty calculated from simulations. On the contrary, in the auto case the Fisher uncertainties resemble those computed from simulations and, they are smaller than the cross simulation uncertainties.  

We have studied the same sky but scaling the noise by a factor of 25 (third row of figure~\ref{fig:comparison_total_cross_auto}). When the noise is the most important contribution instead of the foregrounds the numerical errors are avoided and the results obtained using either the cross or the total information are very similar. The noise increment results in a poorer reconstruction of the polarization angles and larger uncertainties as expected. In this case the information from the auto spectra is very limited and the estimations and uncertainties are noticeably worse than in the cross and total cases. In comparison to the previous scenario, we find that for noise-dominated missions (e.g., Planck) the use of the cross or total information returns superior results. However, in the case of foregrounds-limit missions such as LiteBIRD the auto information retrieves accurate results with the lowest uncertainties. Thus, in the following sections, we limit the study to the auto case. Besides, this is the reason why only auto spectra are used in the previous subsections as well as in \cite{RAC}. 

In addition, we have also analyzed two different skies, one where the foregrounds are Gaussian and anisotropic, and the other Gaussian and isotropic (first and third row of  figure~\ref{fig:comparison_total_cross_auto_sim} respectively). These cases were studied in order to understand the mismatch between the Fisher and simulations uncertainties in the cross case for the CMB+\texttt{s0d0}+N scenario. In the cross case, we find that this discrepancy arises due to the anisotropic nature of the foregrounds as it appears in the CMB+\texttt{AG}+N unlike in the CMB+\texttt{IG}+N. Besides, we observe that the total case is no longer biased. Therefore, we infer that the foregrounds anisotropy and non-gaussianity introduces some discrepancies in the recovered rotation angles since we are solving a Gaussian likelihood. 

Finally, we have studied whether the assumption of having a zero $EB$ foreground contribution could bias our results. This is performed by comparing the results of  CMB+\texttt{AG}+N (CMB+\texttt{IG}+N) with those obtained using the CMB+\texttt{AGnEB}+N (CMB+\texttt{IGnEB}+N) skies. We find that the bias in the auto case is significantly reduced when there is no $EB$ contribution in the simulated foregrounds, while the total and cross cases remain mostly the same. Thus, if the foregrounds $EB$ is not negligible and not taken into account in the likelihood model, the auto solutions might be biased. However, as stated before, current measurements are compatible with a statistically zero foregrounds $EB$ contribution. On the other hand, the removal of the $EB$ foregrounds contribution does not impact the recovered Fisher and simulations uncertainty. This is a result of the contribution to the covariance matrix from the $EB$ power spectrum terms being comparatively lower than those coming from the $EE$ and $BB$ power spectra.

\section{Component Separation}
\label{sec:component_separation}
The characterization of in-flight miscalibration angles is crucial to avoid systematics in CMB component separation analysis. Thus, one can use the results obtained with the above methodology to remove any possible bias introduced by a non-zero  rotation angle. Here, we propose two different approaches to perform component separation adopting the information supplied by the above method. The component separation method B-SeCRET (Bayesian-Separation of Components and Residuals Estimate Tool) applied in this study is an improved version 
of the parametric Bayesian pixel-based  methodology presented in \cite{de2020detection}. The two approaches are the following:
\begin{itemize}
    \item \textbf{Approach 1.} Component separation is applied to the signal maps after being de-rotated using the estimation of rotation angles. The parametric methodology is applied to estimate both the CMB and the foregrounds model parameters.
    \item \textbf{Approach 2.} Component separation is applied to the rotated signal maps. Thus, the rotation angles are included in the model parameters set. We include a Gaussian prior to help determine the rotation angles parameters. The prior's mean is fixed to the rotation angle estimate and the covariance matrix is the inverse of the calculated Fisher matrix obtained with auto-spectra.
\end{itemize}
It is relevant to highlight that the Approach 2 is also a map-based method to estimate the polarization rotation angles as opposed to the power spectrum-based method previously introduced.

 This section is organized as follows. We describe the two component separation approaches used in section~\ref{subsec:cs_approaches}. We compare the polarization angle estimates from Approach 2 and from the power spectrum method in section~\ref{subsec:cs_angles}. Finally, we contrast the CMB recovered using both approaches in section~\ref{subsec:cs_comparison}. 

\subsection{Component Separation Approaches}
\label{subsec:cs_approaches}

Here, we describe the two approaches used to perform component separation in section~\ref{subsubsec:approach1} and section~\ref{subsubsec:approach2}. Both approaches are based on the parametric pixel-based method, B-SeCRET. Since we use a pixel-based method, a large computational time is needed to fit the simulated $N_\mathrm{side}=512$ maps. Thus, we have decided to downgrade the maps to $N_\mathrm{side}=64$ through spherical harmonics with a final resolution of 132 arc-minutes to perform the component separation analysis. 


\subsubsection{Approach 1}
\label{subsubsec:approach1}
B-SeCRET is a Bayesian full-parametric pixel-based maximum likelihood method, which employs an affine-invariant ensemble sampler for Markov Chain Monte Carlo (MCMC) \cite{foreman2013emcee}, to obtain the model parameters. Here we apply the methodology presented in \cite{de2020detection} with the following modifications:
\begin{itemize}
    \item Let $\theta$ be the set of model parameters. $\theta$ is the union of the disjoint sets of: 1) amplitudes parameters \amplitudes\,  and, ii) spectral parameters \spectrals. Instead of fitting both \amplitudes\, and \spectrals\, at the same time, each set is fitted separately. First, the \spectrals\, parameters are fixed to some initial values\footnote{The initial values for the spectral parameters used are the most-likely values given in \cite{akrami2018planck}.} and the \amplitudes\, parameters are calculated. Then, the \amplitudes\, are fixed to the newly obtained values and the \spectrals\, are fitted. This process is repeated iteratively until convergence is reached.
    \item Unlike the \amplitudes\, parameters which vary pixel-wise, the \spectrals\, parameters can have a constant value within a sky region (i.e., clusters of pixels). Thus, we can obtain the value of \spectrals\, in a given cluster of pixels instead of in a pixel. This procedure reduces significantly the statistical uncertainties of the methodology. Since the foregrounds are simulated using the \texttt{s0d0} \texttt{PySM} model, that assumed constant spectral parameters over the sky, we have considered all the available sky as a cluster of pixels\footnote{The available sky considered is the one left by the galactic Planck mask with $f_{sky}=0.60$, (\texttt{HFI\_Mask\_GalPlane-apo0\_2048\_R2.00.fits} downloaded from  \url{https://pla.esac.esa.int/##maps}\cite{ade2014planck}).}.
\end{itemize}
The parametric models, $Q(\theta)$ and $U(\theta)$, that describe the $Q$ and $U$ signals at the pixel $p$ are given by
\begin{equation}
    \begin{pmatrix}
		Q(\theta) \\
	    U(\theta)
	\end{pmatrix}_{p}
	=
	\begin{pmatrix}
		c^{\notsotiny{Q}} \\
	    c^{\notsotiny{U}}
	\end{pmatrix}_{p}
	+ 
	\begin{pmatrix}
		a_s^{\notsotiny{Q}} \\
	    a_s^{\notsotiny{U}}
	\end{pmatrix}_{p}
	\dfrac{1}{u(\nu)}\mleft(\dfrac{\nu}{\nu_s}\mright)^{\beta_{s}} 
	+ 
	\begin{pmatrix}
		a_d^{\notsotiny{Q}} \\
	    a_d^{\notsotiny{U}}
	\end{pmatrix}_{p}
	\dfrac{1}{u(\nu)}
	\mleft(\dfrac{\nu}{\nu_d}\mright)^{\beta_{d}-2}
    \dfrac{B\mleft(\nu,T_{d}\mright)}{B\mleft(\nu_d,T_{d}\mright)}
    \, ,
    \label{eq:cs_parametric_model}
\end{equation}
where $\mathcal{A} = \{c^{\notsotiny{Q}},c^{\notsotiny{U}},a_s^{\notsotiny{Q}},a_s^{\notsotiny{U}},a_d^{\notsotiny{Q}},a_d^{\notsotiny{U}}\}$ correspond to the Q and U amplitudes of the CMB, synchrotron and dust respectively, $\mathcal{B} = \{\beta_s,\beta_d,T_d\}$ are the spectral parameters, $u(\nu) = x(\nu)^2\exp(x(\nu))/(\exp(x(\nu))-1)^2$ and $x =h\nu/(KT_{CMB})$ is a unit conversion factor from thermodynamic to antenna units, $\nu_s = \nu_d = 150$ GHz are the synchrotron and dust pivot frequencies and $B(\nu,T)$ is Planck's law. The second (third) term on the RHS models the synchrotron (dust) contribution. Besides, we have applied Gaussian priors to the spectral parameters: $\beta_{s}\sim \mathcal{N}(-3.1,0.3)$, $\beta_{d}\sim \mathcal{N}(1.56,0.1)$, and $T_{d}\sim \mathcal{N}(21,3)$. 


\subsubsection{Approach 2}
\label{subsubsec:approach2}

In this approach the rotation angles $\alpha$ are included as models parameters as follows
\begin{equation}
    \begin{pmatrix}
		Q \\
	    U
	\end{pmatrix}_{p}
	(\alpha,\theta)
	=
	\begin{pmatrix}
		\cos(2\alpha) & -\sin(2\alpha)\\
	    \sin(2\alpha) & \cos(2\alpha)\\
	\end{pmatrix}
	\begin{pmatrix}
		Q(\theta) \\
	    U(\theta)
	\end{pmatrix}_{p} \, .
\end{equation}
Now, the model parameters are split in three categories, \amplitudes, \spectrals\, and the rotation angles. As in Approach 1, the parameters are calculated iteratively:
\begin{enumerate}
    \item First, the rotation angles $\alpha$ are set to the expected value of the prior, and the \amplitudes\, and \spectrals\, parameters are calculated as in Approach 1. This step is equivalent to the Approach 1 if the expected value of the prior is equal to the rotation angles estimates.
    \item Then, the sky signal parameters are fixed to the ones obtained in the previous step and the rotation angles are fitted. 
    \item The rotation angles are fixed to the results from the last step and this process is repeated until we reach convergence.
\end{enumerate}
One can apply this procedure without using the prior information from the power spectrum based method. In that case, this approach is an independent method to estimate the polarization rotation angles. However, the use of priors helps significantly by speeding up the convergence of the MCMC. Thus, if there is no prior information regarding the calibration of the instrument, the results from the power spectrum based method can be used as prior information.


\subsection{Polarization Rotation Angles Estimation Comparison}
\label{subsec:cs_angles}

\begin{figure}
    \centering
    \includegraphics[width=\linewidth]{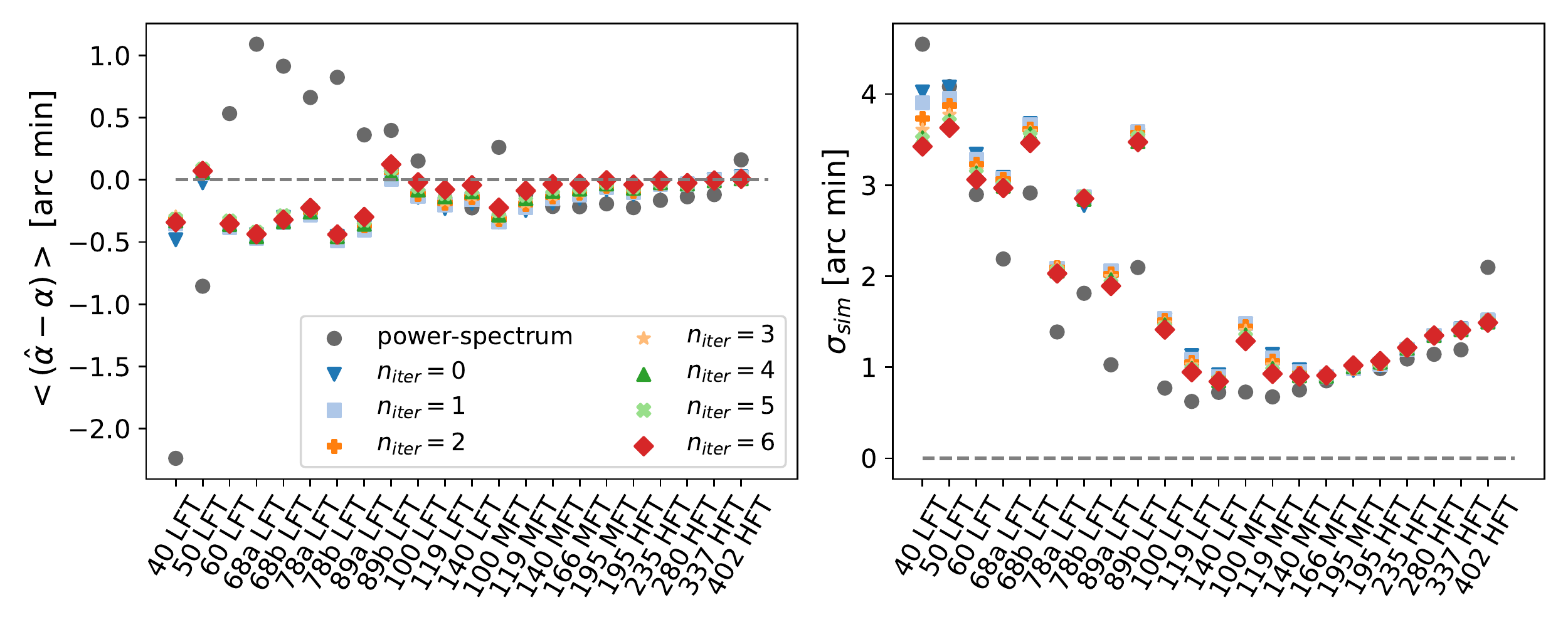}
    \caption{Left panel: Mean difference between the estimated and the actual rotation angle per channel obtained using the power spectrum method and the Approach 2. Right panel: $\sigma_{\mathrm{sim}}$ uncertainty obtained as the simulations dispersion. Since the Approach 2 is an iterative approach, the results from several iterations are displayed. }
    \label{fig:comparison_a2_ps_angles}
\end{figure}

In this section we compare the rotation angle estimates obtained using the Approach 2 to those obtained with the power spectrum based method presented in section~\ref{sec:methodology}. For this purpose, we have applied the Approach 2 to the 100 rotated signal maps simulations. Figure~\ref{fig:comparison_a2_ps_angles} shows results using both methods.  

We observe that the bias of the rotation angles estimates is reduced after applying Approach 2. These differences are further enhanced as the number of iterations increases. Besides, the right panel of figure~\ref{fig:comparison_a2_ps_angles} shows that the uncertainty of the rotation angles estimates obtained from simulations in the Approach 2 also decreases with the number of iteration. On the other hand, the uncertainties obtained with the power spectrum based method are slightly better than those recovered with Approach 2.


It is worth noting that, the rotation angles in Approach 2 are estimated using 60\% of the sky unlike in the power spectrum method where the full sky was used in the analysis. On the one hand, larger sky fractions will lead to a reduction on the uncertainty of the rotation angles obtained with Approach 2. However, the inclusion of Galactic plane regions might yield biased estimates of the rotation angles due to incorrect foreground modelling. Moreover, the Approach 2 is computationally more expensive than the Approach 1 since there is an additional step involved in the fitting procedure. On the other hand, real data analyses of rotation angles with the power spectrum method will require at least a small mask to hide point sources and CO bright regions. Therefore, power spectra estimators algorithm's will be needed in order to obtain the true power spectra of the maps. This issue introduces more uncertainty on the rotation angles estimates from the power spectrum model. Hence both methods could be conceived as complementary and an useful consistency test.

\subsection{CMB Recovery Comparison}
\label{subsec:cs_comparison}

\begin{figure}
    \begin{subfigure}{.5\linewidth}
         \centering
        \includegraphics[width=\linewidth]{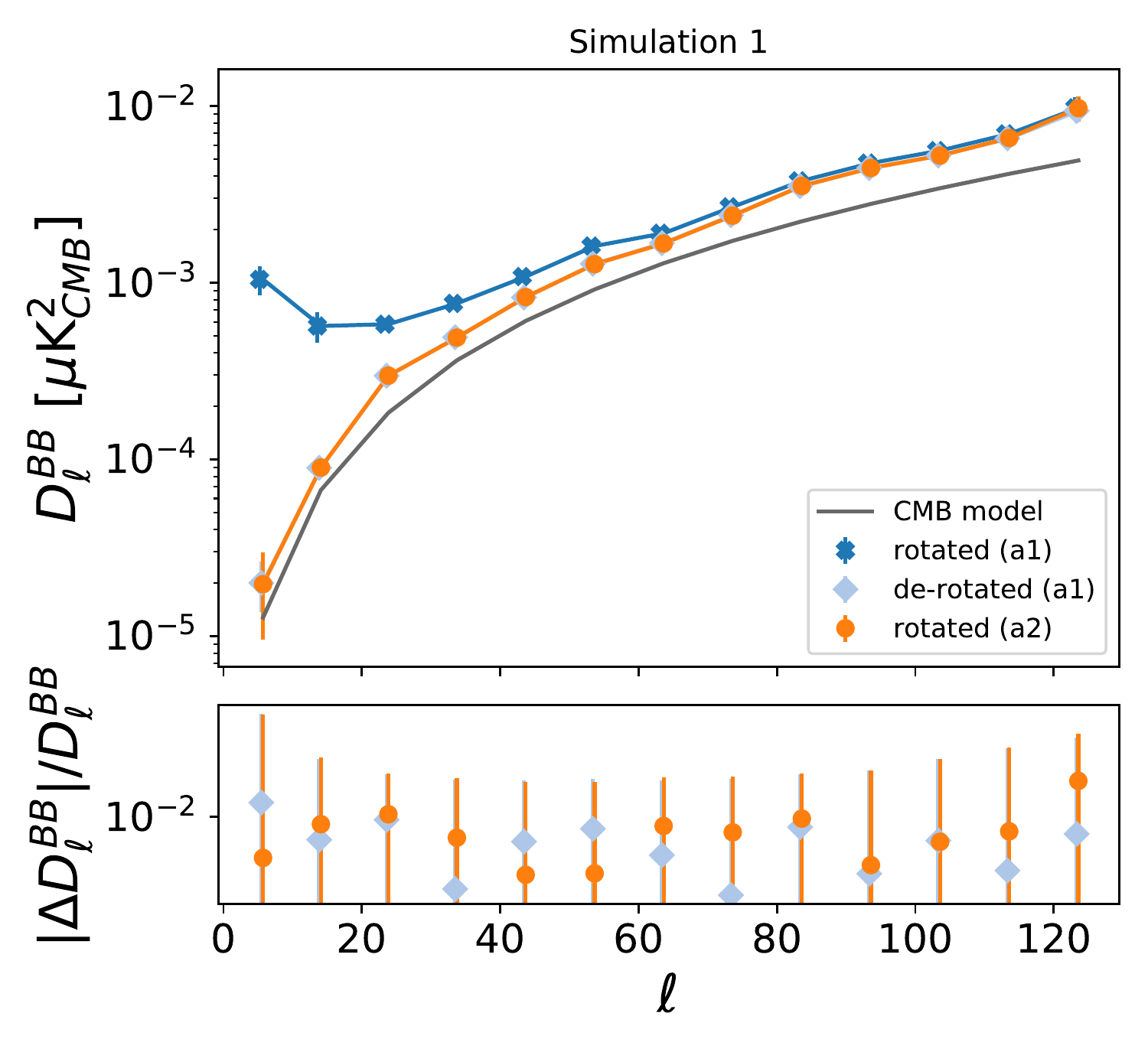}
    \end{subfigure}  
    \begin{subfigure}{.5\linewidth}
        \centering
        \includegraphics[width=\linewidth]{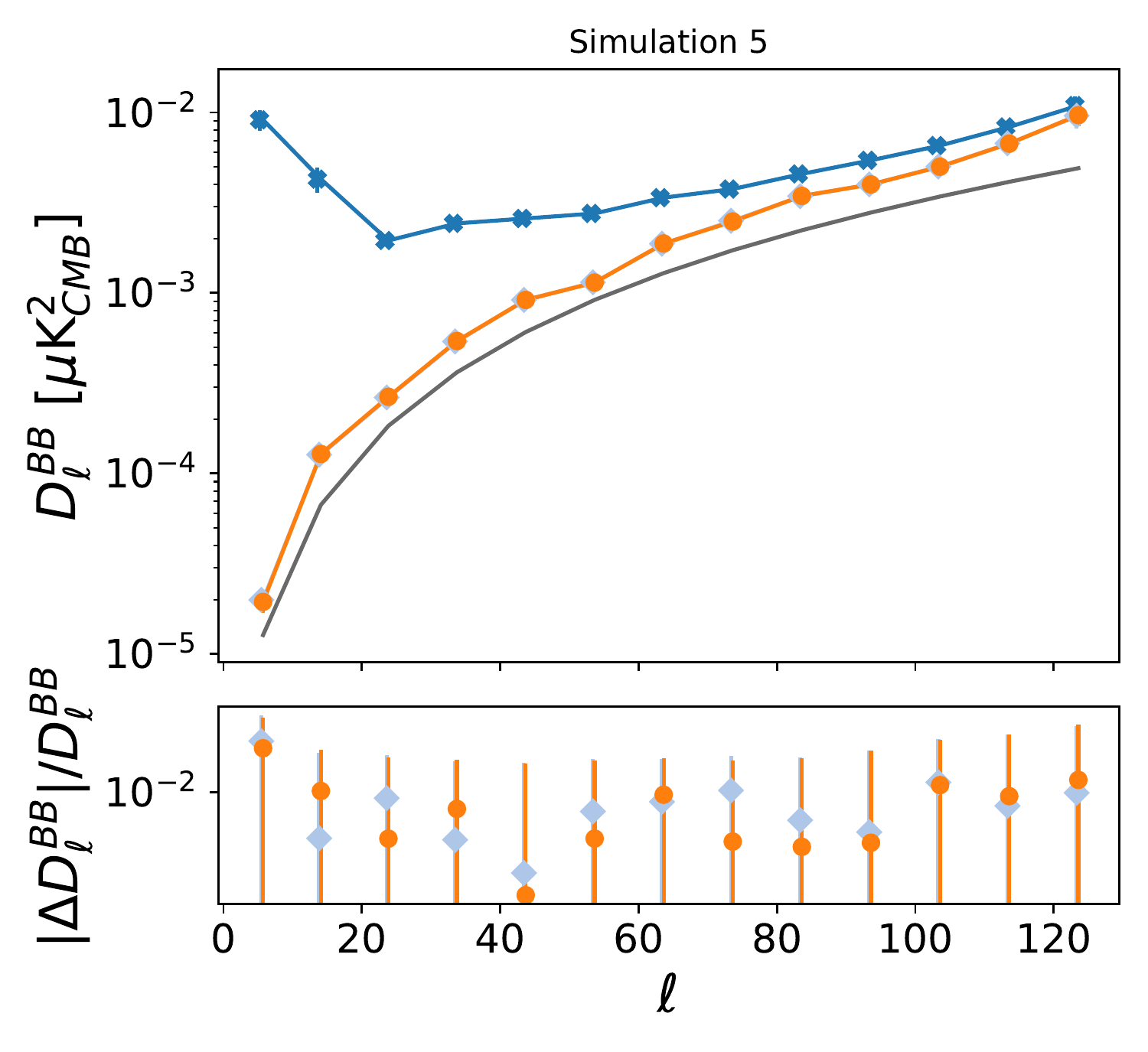}
    \end{subfigure}
    \caption{$B$-power spectrum from cleaned CMB maps obtained using the Approach 2 (orange circles) and  the Approach 1 on the rotated (dark blue stars) and the de-rotated maps (light blue diamonds). The bottom panel displays the absolute relative residuals of the de-rotated (a1) as well as rotated (a2) with respect to the non-rotated case. The black line corresponds to the CMB $B$-power spectrum model. The difference between the model and the recovered power spectrum appears due to a non-removal of the noise bias.}
    \label{fig:comparison_ps_approaches}
\end{figure}

In this section, we compare the CMB maps recovered  with both approaches, i.e., the CMB obtained after applying the Approach 1 to the de-rotated maps and the Approach 2 to the maps with this systematic (rotated maps). Moreover, we have applied the Approach 1 to the maps without the systematic (non-rotated maps), and the rotated map to study the improvement on the CMB cleaning after mitigating the effect of non-null polarization rotation angles. The angular power spectra is calculated using \texttt{ECLIPSE}, a fast Quadratic Maximum Likelihood estimator developed by \cite{eclipse_qml}. The results from the $1^{\rm{st}}$ and $5^{\rm{th}}$ simulation are shown in figure~\ref{fig:comparison_ps_approaches}.

From figure~\ref{fig:comparison_ps_approaches} it is clear that the worst results correspond to the Approach 1 applied to the rotated maps as expected since in this case we do not correct the effect induced by non-zero polarization angles. In order to study the systematic error left on the cleaned CMB maps, we show in the bottom panel the absolute relative residuals of the Approach 1 applied to the de-rotated maps as well as the Approach 2 applied to the rotated maps with respect to the non-rotated maps. The residuals of cleaned CMB maps using both Approach 1 and Approach 2 are  of the order of a 1\% and are compatible between them. Although the Approach 2 might provide a better estimation of the rotation angles, both approaches return similar systematic residuals. The rotation angle estimates from the power spectrum method are sufficiently accurate so as to mitigate most of the systematic induced in the CMB. The slight improvement in the angle estimation by Approach 2 does not lead to a better cleaning of the CMB. The systematic residuals left do not bias the tensor-to-scalar ratio $r$ recovered from the cleaned CMB maps as shown in \cite{RAC}.

\section{Conclusions}
\label{sec:conclusions}
We have introduced an iterative angular power spectra maximum likelihood-based method to calculate the polarization rotation angles from the multi-frequency signal by nulling the $EB$ power spectrum. Two major assumptions are made: i) the rotation angles are small ($\leq6^\circ$), and, ii) the covariance matrix does not depend on the rotation angles. Under this framework, we obtain analytical equations for both the angles and their uncertainties, which leads to a very fast computational method.

We have studied the performance of this methodology using sky simulations of a LiteBIRD-like experiment. These studies demonstrate that binning the observed spectra seems to be enough to overcome the lack of a reliable model for the angular power spectra of Galactic foregrounds for the calculation of the covariance matrix, and that, for the typical noise levels expected for next-generation experiments, the methodology will provide an unbiased estimation of rotation angles with a few arcminutes precision. The comparison of the full-instrument versus a telescope-by-telescope analysis also showed the power of our multi-frequency approach, since the overall uncertainties improve when information from different bands are combined.

Although the formalism is capable of exploiting all combinations of $EB$ cross-spectra in combination with auto-spectra, we found that rotation angles estimated in this case are biased for typical sky simulations. This discrepancy comes from numerical errors that arise from approximating the theoretical power spectra in the covariance matrix calculation with the actual observed power spectra. We showed that these numerical instabilities disappear when the noise is larger than the foregrounds contribution. We found that for noise-dominated experiments it is better to use all available information (i.e., cross-spectra and auto-spectra) whereas, for foreground-dominated experiments such as LiteBIRD, limiting the information to only the auto-spectra provides results with smaller uncertainties. 

Moreover, we have proposed two different approaches with which the rotation angles results from this methodology can be employed in the component separation analysis to remove any systematic error in the CMB cleaning. In Approach 1 we apply the B-SeCRET methodology to the signal maps de-rotated with the rotation angle estimates. In Approach 2, we include the rotation angles as model parameters and use the results from the power spectrum methodology as prior information. The Approach 2 can also be used as an independent method to calculate the rotation angles. However, the prior information from the power spectrum method helps significantly with the MCMC convergence. We found that the Approach 2 improves the accuracy in the estimation of the rotation angles, but returns slightly higher uncertainties. Finally, we have compared the CMB recovered using both Approach 1 and Approach 2 with the cleaned CMB obtained from signal maps without any rotation angles. We found that the residuals from both approaches are compatible and of  the order of the 1\% at the power spectrum level.

\acknowledgments

EdlH acknowledges  financial support from the \textit{Concepci\'on Arenal Programme} of the Universidad de Cantabria. PDP acknowledges  financial support from the \textit{Formaci\'on del Profesorado Universitario (FPU) programme} of the Spanish Ministerio de Ciencia, Innovaci\'on y Universidades. We acknowledge Santander Supercomputaci\'on support group at the Universidad de Cantabria, member of the Spanish Supercomputing Network, who provided access to the Altamira Supercomputer at the Instituto de F\'isica de Cantabria (IFCA-CSIC) for performing simulations/analyses. The authors would like to thank the Spanish Agencia Estatal de Investigaci\'on (AEI, MICIU) for the financial support provided under the projects with references PID2019-110610RB-C21, ESP2017-83921-C2-1-R and AYA2017-90675-REDC, co-funded with EU FEDER funds, and also acknowledge the funding from Unidad de Excelencia Mar{\'\i}a de Maeztu (MDM-2017-0765). Some of the results in this paper have been derived using \texttt{CAMB} \cite{lewis2011camb}, \texttt{ECLIPSE} \cite{eclipse_qml}, and the \texttt{healpy} \cite{gorski2005healpix}, \texttt{emcee} \cite{foreman2013emcee}, \texttt{PySM} \cite{thorne2017python}, and \texttt{matplotlib} \cite{matplotlib} \texttt{Python} packages.

\bibliographystyle{JHEP}
\bibliography{references}

\end{document}